\documentclass[12pt,a4paper]{article}

\usepackage{amsmath,amssymb}
\usepackage[nosort]{cite}
\usepackage[hyperref,bulletsep]{collect}

\makeatletter \@addtoreset{equation}{section} \makeatother

\makeatletter
\let\old@makecaption=\@makecaption
\def\@makecaption{\small\old@makecaption}
\makeatother

\makeatletter
\let\old@startsection=\@startsection
\renewcommand{\@startsection}[6]{\old@startsection{#1}{#2}{#3}{#4}{#5}{#6\mathversion{bold}}}
\makeatother

\let\oldPhi=\Phi
\let\oldPsi=\Psi
\let\oldGamma=\Gamma
\let\oldLambda=\Lambda
\let\oldDelta=\Delta
\let\oldSigma=\Sigma
\let\oldTheta=\Theta
\let\oldPi=\Pi
\renewcommand{\Phi}{\mathnormal{\oldPhi}}
\renewcommand{\Psi}{\mathnormal{\oldPsi}}
\renewcommand{\Gamma}{\mathnormal{\oldGamma}}
\renewcommand{\Sigma}{\mathnormal{\oldSigma}}
\renewcommand{\Delta}{\mathnormal{\oldDelta}}
\renewcommand{\Theta}{\mathnormal{\oldTheta}}
\renewcommand{\Lambda}{\mathnormal{\oldLambda}}
\renewcommand{\Pi}{\mathnormal{\oldPi}}

\newcommand{\smat}{\mathcal{S}}

\newcommand{\superN}{\mathcal{N}}

\newcommand{\sign}{\mathop{\mathrm{sign}}}
\newcommand{\Li}{\mathop{\mathrm{Li}}\nolimits}

\newcommand{\order}[1]{\mathcal{O}(#1)}

\newcommand{\Integers}{\mathbb{Z}}
\newcommand{\Reals}{\mathbb{R}}
\newcommand{\Comp}{\mathbb{C}}

\ifx\genfrac\sdflkaj

\else

\fi
\newcommand{\sfrac}[2]{{\textstyle\frac{#1}{#2}}}
\newcommand{\half}{\sfrac{1}{2}}

\newcommand{\quarter}{\sfrac{1}{4}}

\newcommand{\supup}[1]{^{\mathrm{#1}}}

\newcommand{\alg}[1]{\mathfrak{#1}}

\newcommand{\lrbrk}[1]{\left(#1\right)}
\newcommand{\bigbrk}[1]{\bigl(#1\bigr)}
\newcommand{\brk}[1]{(#1)}

\newcommand{\xvar}{x}
\newcommand{\xexp}[2]{\xvar^{(#1)}_{#2}}
\newcommand{\xexpb}[2]{\xvar^{#1}_{#2}}
\newcommand{\xp}[1]{\xvar^+_{#1}}
\newcommand{\xm}[1]{\xvar^-_{#1}}
\newcommand{\xpm}[1]{\xvar^\pm_{#1}}
\newcommand{\xmp}[1]{\xvar^{\mp}_{#1}}
\newcommand{\xpmbar}[1]{\bar{\xvar}^\pm_{#1}}
\newcommand{\xpmbarbar}[1]{\bar{\bar{\xvar}}^\pm_{#1}}

\makeatletter
\newcommand{\ellSN}{\mathop{\operator@font sn}\nolimits}
\newcommand{\ellCN}{\mathop{\operator@font cn}\nolimits}
\newcommand{\ellDN}{\mathop{\operator@font dn}\nolimits}
\newcommand{\ellAM}{\mathop{\operator@font am}\nolimits}
\newcommand{\ellK}{\mathop{\smash{\operator@font K}\vphantom{a}}\nolimits}
\newcommand{\ellE}{\mathop{\smash{\operator@font E}\vphantom{a}}\nolimits}
\newcommand{\gammafn}{\mathop{\smash{\oldGamma}\vphantom{a}}\nolimits}
\newcommand{\bernoulli}{\mathrm{B}}
\renewcommand{\digamma}{\mathop{\smash{\oldPsi}\vphantom{a}}\nolimits}
\renewcommand{\Re}{\mathop{\operator@font Re}\nolimits}
\renewcommand{\Im}{\mathop{\operator@font Im}\nolimits}
\makeatother

\newcommand{\nln}{\nonumber\\}
\newcommand{\nl}[1][0pt]{\nonumber\\[#1]&\hspace{-4\arraycolsep}&\mathord{}}
\newcommand{\nlnum}{\\&\hspace{-4\arraycolsep}&\mathord{}}
\newcommand{\earel}[1]{\mathrel{}&\hspace{-2\arraycolsep}#1\hspace{-2\arraycolsep}&\mathrel{}}
\newcommand{\eq}{\earel{=}}

\def\[{\begin{equation}}
\def\]{\end{equation}}
\def\<{\begin{eqnarray}}
\def\>{\end{eqnarray}}
\def\be{\begin{equation}}
\def\ee{\end{equation}}
\def\ba{\begin{eqnarray}}
\def\ea{\end{eqnarray}}

\makeatletter
\def\mr@ignsp#1 {\ifx\:#1\@empty\else #1\expandafter\mr@ignsp\fi}%
\newcommand{\multiref}[1]{\begingroup
\xdef\mr@no@sparg{\expandafter\mr@ignsp#1 \: }%
\def\mr@comma{}%
\@for\mr@refs:=\mr@no@sparg\do{\mr@comma\def\mr@comma{,}\ref{\mr@refs}}%
\endgroup}
\makeatother

\newcommand{\hypref}[2]{\ifx\href\asklfhas #2\else\href{#1}{#2}\fi}

\newcommand{\appref}[1]{App.~\multiref{#1}}

\renewcommand{\eqref}[1]{(\multiref{#1})}

\ifx\href\asklfhas\newcommand{\href}[2]{#2}\fi
\newcommand{\arxivno}[1]{\href{http://arxiv.org/abs/#1}{#1}}

\begin{document}

\thispagestyle{empty}
\begin{flushright}\footnotesize
\texttt{\arxivno{hep-th/0609044}}\\
\texttt{AEI-2006-068}\\
\texttt{CERN-PH-TH/2006-176}\\
\texttt{IFT-UAM/CSIC-06-44}\\
\texttt{PUTP-2208}%
\vspace{0.5cm}
\end{flushright}
\vspace{0.1cm}

\renewcommand{\thefootnote}{\fnsymbol{footnote}}
\setcounter{footnote}{0}

\begin{center}%
{\Large\textbf{\mathversion{bold}%
A Crossing--Symmetric Phase\\
for $AdS_5 \times S^5$ Strings}\par}
\vspace{1cm}

\textbf{Niklas Beisert$^{1,2}$, Rafael Hern\'andez$^3$} and \textbf{Esperanza L\'opez$^4$}
\vspace{5mm}

\textit{${}^1\!$
Max-Planck-Institut f\"ur Gravitationsphysik,
Albert-Einstein-Institut\\
Am M\"uhlenberg 1, 14476 Potsdam, Germany} \vspace{3mm}

\textit{${}^2\!$
Physics Department, Princeton University\\
Princeton, NJ 08544, USA} \vspace{3mm}

\textit{${}^3\!$ Theory Division, CERN\\
CH-1211 Geneva 23, Switzerland} \vspace{3mm}

\textit{${}^4\!$ Instituto de F\'{\i}sica Te\'orica, UAM-CSIC\\
Cantoblanco, 28049 Madrid, Spain} \vspace{3mm}

\vspace{1.5cm}

\textbf{Abstract}\vspace{5mm}

\begin{minipage}{14.8cm}
We propose an all-order perturbative 
expression for the dressing phase of the 
$AdS_5\times S^5$ string S-matrix at strong coupling.
Moreover, we are able to sum up large parts of this expression.
This allows us to start the investigation of the analytic structure 
of the phase at finite coupling revealing a few surprising features. 
The phase obeys all known constraints including the crossing relation and it
matches with the known physical data at strong coupling. In particular,
we recover the bound states of giant magnons recently found by
Hofman and Maldacena as poles of the scattering matrix.
At weak coupling our 
proposal seems to differ with gauge theory. 
A possible solution to this disagreement is the inclusion of additional 
pieces in the phase not contributing to crossing, which we also study.

\end{minipage}

\vspace*{\fill}

\end{center}

\hrule
\vspace{1.5mm}

\texttt{\footnotesize{nbeisert@aei.mpg.de, rafael.hernandez@cern.ch, esperanza.lopez@uam.es}}
\par\vspace{-.4cm}

\newpage
\setcounter{page}{1}
\renewcommand{\thefootnote}{\arabic{footnote}}
\setcounter{footnote}{0}

\section{Introduction}

The quantum description of type IIB string theory on $AdS_5\times S^5$
remains a challenge because quantisation of the Metsaev-Tseytlin action
\cite{Metsaev:1998it} in the conformal gauge faces a number of intricate problems.
Insight into a way to overcome this obstacle has arisen along the last years
from the observation that the classical sigma model for the string on
$AdS_5 \times S^5$ is integrable \cite{Bena:2003wd}. Integrability of the
string implies that it admits a Lax connection, and allows a resolution of the
spectrum of classical strings in terms of spectral curves \cite{Kazakov:2004qf,Beisert:2005bm}.
The integral equations satisfied by the spectral density for the monodromy of 
the Lax connection remind of a thermodynamic limit of some Bethe equations, and a 
set of discrete Bethe equations for the quantum string sigma model were in fact 
suggested in \cite{Arutyunov:2004vx,Beisert:2005fw}. 
Integrable structures also arise on the
gauge theory side of the AdS/CFT correspondence because the leading planar dilatation
operator of $\superN=4$ supersymmetric Yang-Mills has been identified with the
Hamiltonian of an integrable spin chain \cite{Lipatov:1997vu,Minahan:2002ve,Beisert:2003yb}. 
Moreover, integrability has also been shown to hold at
higher loops in some restricted sectors \cite{Beisert:2003tq,Beisert:2003ys,Serban:2004jf,Staudacher:2004tk}.
Assuming that integrability holds, Bethe equations have been
proposed as an efficient means
to describe the spectrum of $\superN=4$ Yang-Mills operators.
For further details and references we would like to refer 
the reader to the reviews \cite{Beisert:2004ry,Plefka:2005bk}.

The asymptotic Bethe ans\"atze for gauge and string theory are very
similar, and the asymptotic S-matrices on each side of the correspondence 
differ simply by a scalar factor \cite{Staudacher:2004tk,Beisert:2005fw}. 
Actually, this is as much as they could possibly
differ: The two-excitation S-matrix of an
infinite spin chain system with the $\alg{psu}(2,2|4)$ symmetry 
that characterises both string theory on $AdS_5\times S^5$ 
and $\superN=4$ super Yang-Mills can be fixed up to a scalar factor \cite{Beisert:2005tm}
\[
\smat_{12}=
S^0_{12} \;
\smat\supup{\alg{su}(2|2)}_{12} \,
\smat\supup{\alg{su}(2|2)\prime}_{12} \ .
\label{smatrix}
\]
The spin chain vacuum breaks the $\alg{psu}(2,2|4)$ symmetry algebra down to
$\alg{psu}(2|2)^2 \ltimes \mathbb{R}$, where $\mathbb{R}$ represents a
shared central charge \cite{Beisert:2004ry}. In order to describe elementary
excitations of the chain, it is necessary to extend the unbroken symmetry
algebra with two central charges. The symbol $\smat\supup{\alg{su}(2|2)}_{12}$
denotes the uniquely fixed flavour structure of the S-matrix for each
centrally extended $\alg{su}(2|2)$ sector.

The determination of the scattering matrix
by the symmetries up to a scalar
factor is not unique to the AdS/CFT chain, but a generic fact in integrable
systems \cite{Zamolodchikov:1978xm}. In order to determine
the dressing factor,
additional dynamical information, such as crossing symmetry for relativistic
systems, is required.
The status of crossing symmetry in the AdS/CFT context is not a priori clear
since the dispersion relation of elementary excitations
does not have precise relativistic invariance.
However it has been argued that
crossing symmetry should still hold for strings on $AdS_5 \times S^5$ \cite{Janik:2006dc}.
A strong argument in favour is that a purely algebraic implementation of
crossing symmetry based on the underlying Hopf algebra structure of
integrable systems, known to work in well studied examples, leads to a
non-trivially consistent picture \cite{Janik:2006dc,Gomez:2006va,Plefka:2006ze}.
Furthermore, it was shown in \cite{Beisert:2005tm} that the constraint from
crossing symmetry is equivalent to a certain bootstrap condition
implying that a particle-hole pair should scatter trivially.
Moreover, the classical string
phase factor \cite{Arutyunov:2004vx} plus its one-loop
string sigma model quantum correction \cite{Hernandez:2006tk} have been shown to satisfy
crossing to the appropriate order \cite{Arutyunov:2006iu}.
Recently, a function
satisfying crossing has been presented in \cite{Beisert:2006zy}.
The aim of this work is a proposal for a general dressing factor which
obeys the crossing relation found in \cite{Janik:2006dc}
and agrees with the available perturbative data from string theory.

The plan of the paper is as follows.
In section \ref{sec:particle}
we will review the description of elementary excitations developed
in \cite{Beisert:2005tm} and the formulation of crossing symmetry \cite{Janik:2006dc}
in order to make our presentation self-contained.
We will end this section with a discussion
of the separate kinematical regimes that characterise the strong and weak
coupling limits.
In section \ref{sec:series} we will describe the dressing phase
of the scattering matrix in terms
of a perturbative series at strong coupling and
propose a concrete expression for the coefficients that govern the series.
These coefficients are a natural extension of those determining the classical
dressing phase \cite{Arutyunov:2004vx} and its one-loop correction \cite{Hernandez:2006tk}. We will
provide evidence that the proposed series satisfies crossing symmetry.
An analytic expression is presented in section \ref{sec:function} and argued to represent
the resummed series. Using this result we are able to identify the bound states
of giant magnons recently found in \cite{Hofman:2006xt} as poles of the scattering
matrix. Although our dressing phase was constructed to satisfy the
main physical requirements on the string side, such as crossing symmetry,
we should stress that it does not correctly connect with gauge theory in the
weak coupling regime.
A possible way out is the addition of a homogeneous
solution of the crossing equation to the dressing phase. The study of homogeneous
solutions is addressed in section \ref{sec:homo}.
In section \ref{sec:concl} we discuss our results
and comment on the many open issues.
The paper concludes with two appendices that collect useful formulae
for the weak and strong coupling expansions.

\section{Particle Model}
\label{sec:particle}

We will start by reviewing the model of physical
excitations above a half-BPS vacuum state.
This section describes the setup as well as the
basic definitions and conventions to be used
in later sections of this paper.

\subsection{Setup}

States in type IIB string theory are naturally described by a
set of 8 bosonic and 8 fermionic excitations propagating on a circle.
For $\superN=4$ gauge theory the setup is similar except
that the circle is replaced by a periodic spin chain \cite{Berenstein:2002jq}.
The vacuum state is a protected half-BPS state in both cases
and each particle has an associated momentum $p_k$.
Due to the compactness of the circle or the spin chain,
the spectrum of states is discrete.
Discreteness is achieved by imposing
quantisation conditions on the particle momenta.

However, it is more convenient to replace the circle with an infinite line.
This relaxes the quantisation condition of particle momenta and makes
the spectrum continuous.
To recover the circle we need to impose periodicity conditions
on the multi-particle wave function,
the so-called Bethe equations.
The Bethe equations rely on the scattering matrix $\smat$
of particles on the infinite line \cite{Staudacher:2004tk}.

The symmetry of the full model
is $\alg{psu}(2,2|4)$ and a subalgebra
$\Reals\ltimes \alg{psu}(2|2)^2\ltimes \Reals$
preserves the particle numbers \cite{Beisert:2004ry}.
The external automorphism is the $\alg{so}(6)$ charge $J$ and 
the central charge $C$ measures the energy of a state.
On the infinite line, the residual algebra enlarges by two central charges
$\Reals\ltimes \alg{psu}(2|2)^2\ltimes \Reals^3$ \cite{Beisert:2005tm}.
These central charges describe the momentum of a particle.

\subsection{Particles}

A particle is described by its momentum $p$, energy $C$
(alias the $\alg{su}(2|2)$ central charge)
and its flavour.
There are sixteen particle flavours which form
a multiplet of the residual symmetry.
The momentum and energy are related by the dispersion relation \cite{Beisert:2004hm}
\[\label{eq:shell}
4C^2-16 g^2 \sin^2(\half p) = 1\ ,
\]
where $g$ is proportional to the square root of the 't Hooft coupling constant,
\[
g=\frac{\sqrt{\lambda}}{4 \pi} \ .
\]
This dispersion relation is in fact an atypicality condition for a short multiplet
of the residual algebra \cite{Beisert:2005tm} and thus appears
to be protected from quantum corrections.

This equation is neither a standard lattice nor a standard relativistic
dispersion relation, but it shares features of both:
It is periodic in the momentum $p$, i.e.~it has the Brillouin zones
of a discrete system. It is also relativistic if we consider
$\sin (\half p)$ (rather than $p$) to be the relevant relativistic momentum.
These two properties square nicely with the observation
that the kinematic space of the elementary excitations defines a complex torus
\cite{Janik:2006dc}.
The torus has two non-trivial cycles, let us call them ``real''
and ``imaginary''.
The real cycle corresponds to periodicity of the momentum $p$
for a lattice model.
The imaginary cycle corresponds to periodicity
of the mass shell condition \eqref{eq:shell} for imaginary relativistic momentum,
i.e.~$(2C)^2+(4ig\sin(\half p))^2=1$ defines a unit circle.

We will use complex variables $\xpm{}$ to codify the momentum $p$
and energy $C$ of physical excitations via
\[
e^{ip}=\frac{\xp{}}{\xm{}}\ ,\qquad
C=\frac{1}{2}+\frac{ig}{\xp{}}-\frac{ig}{\xm{}}\ .
\]
These two variables are subject to the constraint
\[
\xp{} + \frac{1}{\xp{}}-\xm{} -\frac{1}{\xm{}}=\frac{i}{g} \ ,
\label{def}
\]
which is equivalent to the dispersion relation \eqref{eq:shell}.
Furthermore we would like to introduce the auxiliary variable $u$ as
\[\label{udef}
u=\xp{}+\frac{1}{\xp{}}-\frac{i}{2g}=\xm{}+\frac{1}{\xm{}}+\frac{i}{2g}\ .
\]

Finally, we shall present two relevant discrete symmetries.
One of them is parity which maps the particle variables as follows
\[\label{paritymap}
\xpm{}\mapsto -\xmp{},
\qquad
p\mapsto -p,
\qquad
C\mapsto C,
\qquad
u\mapsto -u \ .
\]
For the definition of crossing symmetry
we will furthermore need
the antipode map between particles and particle-holes
given by
\[\label{crossmap}
\xpm{}\mapsto \frac{1}{\xpm{}},
\qquad
p\mapsto -p,
\qquad
C\mapsto -C,
\qquad
u\mapsto u.
\]
%

\subsection{Scattering and Crossing}

The pairwise scattering matrix $\smat_{12}$ for the above particles
was derived in \cite{Beisert:2005tm}. It is fully constrained by symmetry
up to a scalar prefactor $S_{12}^0$.
We will not need its full form here, but only consider the scalar factor.
It will be convenient for us to use the definition
in terms of the dressing factor 
$\sigma_{12}=\sigma(\xpm{1},\xpm{2})$
and dressing phase $\theta_{12}$ introduced in \cite{Beisert:2005fw}
\[
S^0_{12}=
\brk{\sigma_{12}}^2 \,
\frac{\xp{1}-\xm{2}}{\xm{1}-\xp{2}}\,
\frac{1-1/\xm{1}\xp{2}}{1-1/\xp{1}\xm{2}}\,,
\qquad
\sigma_{12}=\exp(i\theta_{12}) \ .
\label{sphase}
\]
The aim of the present paper is to derive
an expression for the dressing phase $\theta_{12}$
consistent with string theory.
Note that in our conventions the dressing phase
appears with a factor of $+2i$ in the exponent of the 
$\alg{psu}(2,2|4)$ scattering matrix
\[
\smat_{12}\sim \exp(2i\theta_{12})\ .
\]

A constraint on the form of $\sigma_{12}$ is gained
by imposing crossing symmetry \cite{Janik:2006dc}.
Crossing symmetry relates the two-particle S-matrix
with the S-matrix for a particle and a particle-hole.
Assuming crossing symmetry holds,
the following constraint on the dressing factor
$\sigma_{12}=\sigma(\xpm{1},\xpm{2})$ was derived in \cite{Janik:2006dc},
\[
\sigma_{12}\,\sigma_{\bar 12} = h_{12} \ ,
\label{cross}
\]
where the bar stands for the replacement of a particle
by a particle-hole, cf.~\eqref{crossmap}.
The function $h$ is given by \cite{Arutyunov:2006iu}
\[\label{h}
h_{12}=\frac{\xm{2}}{\xp{2}} \, \frac{\xm{1} - \xp{2} }{ \xp{1} - \xp{2}} \,
\frac{1-1/\xm{1}\xm{2}}{ 1-{1/ \xp{1} \xm{2}}} \ .
\]

The crossing relation \eqref{cross} looks superficially puzzling because the
l.h.s.~is naively
symmetric under the particle-hole interchange, while the r.h.s.~is not.
It was shown in \cite{Janik:2006dc} that the operation $\xpm{}\mapsto \xpmbar{}=1/\xpm{}$
corresponds to a displacement on the imaginary cycle
of the complex torus by half a period.
Therefore applying twice this operation we move once around a non-trivial
cycle of the torus, which can result in a non-trivial monodromy,
i.e.~$\xpm{}\mapsto \xpmbar{} \mapsto \xpmbarbar{}$.
This is indeed the case since \eqref{h} implies
\[
\sigma_{\bar{\bar{1}}2}=\frac{h_{\bar 12}}{h_{12}}\,\sigma_{12} \neq  \sigma_{12}\, .
\label{dcross}
\]
The operation $\xpm{}\mapsto \xpmbarbar{}$,
which is superficially the identity map,
has the interpretation of a change in
Riemann sheet for the function $\sigma_{12}=\sigma(\xpm{1},\xpm{2})$.
It is therefore instructive to split the dressing
factor into an ``odd'', an ``even'' and a ``homogeneous'' part,
$\sigma=\sigma\supup{odd}\sigma\supup{even}\sigma\supup{hom}$,
where the odd part is responsible for generating
the monodromy in the double crossing relation \eqref{dcross}, while
the even part is a homogeneous solution of double crossing.
These factors individually obey the crossing relations
\[
\sigma\supup{odd}_{12}\,\sigma\supup{odd}_{\bar 12} = h\supup{odd}_{12},\qquad
\sigma\supup{even}_{12}\,\sigma\supup{even}_{\bar 12} = h\supup{even}_{12},\qquad
\sigma\supup{hom}_{12}\,\sigma\supup{hom}_{\bar 12} = 1 \ ,
\label{crossoddeven}
\]
with the odd and even parts of the crossing function
\<
\label{hoddeven}
h\supup{odd}_{12}\eq\sqrt{\frac{h_{12}}{h_{\bar 12}}}
=\sqrt{
\frac{\xm{1}-\xp{2}}{\xp{1}-\xp{2}} \,
\frac{\xp{1}-\xm{2}}{\xm{1}-\xm{2}} \,
\frac{1-1/\xp{1}\xp{2}}{1-1/\xm{1}\xp{2}} \,
\frac{1-1/\xm{1}\xm{2}}{1-1/\xp{1} \xm{2}}
}
 \ ,
\nln
h\supup{even}_{12}\eq
\sqrt{h_{12}\,h_{\bar 12}}
=
\frac{\xm{2}}{\xp{2}}\sqrt{\frac{\xm{1}-\xp{2}}{\xp{1}-\xm{2}}
\,\frac{1-1/\xm{1}\xp{2}}{1-1/\xp{1}\xm{2}}}
=
\frac{\xm{2}}{\xp{2}}\sqrt{\frac{u_1-u_2-i/g}{u_1-u_2+i/g}}\ ,
\>
where $u$ is the crossing-invariant variable \eqref{udef}.
Clearly, the relations \eqref{crossoddeven,hoddeven} are
equivalent to the full crossing relation \eqref{cross}.
The general solution of the crossing relation
is not unique and consequently includes a homogeneous part. 
In the absence of further physical
constraints this homogeneous piece can be chosen arbitrarily.

\subsection{Limits}

Before we consider solutions to these equations, we shall
investigate the strong-coupling and weak-coupling regimes.
In these limits, the kinematic space of particles splits up into disconnected regions. 
These regions give rise to different kinds of particles with different properties.
They will play an important role in perturbative representations of the phase.
Here we will only present a list of such regimes. Explicit formulae
can be found in \appref{sec:strongweak}.

At strong coupling, the kinematic space of particles splits up into four
interesting regions. For later convenience we will denote these four regimes
by MT (Metsaev-Tseytlin plane-wave excitations \cite{Metsaev:1998it}),
HM (Hofman-Maldacena regime \cite{Hofman:2006xt})
and GKPr, GKPl (Gubser-Klebanov-Polyakov flat space limit
\cite{Gubser:1998bc} with distinct right and left-movers):
\[
g\to\infty\quad \Longrightarrow \quad
\mbox{particle} \in
\begin{cases}
\mbox{MT} & \mbox{if } p=\order{1/g^1}\ ,
\\
\mbox{GKPr} & \mbox{if } p=\order{1/g^{1/2}}\mbox{ and }p>0\ ,
\\
\mbox{GKPl} & \mbox{if } p=\order{1/g^{1/2}}\mbox{ and }p<0\ ,
\\
\mbox{HM} & \mbox{if } p\in (0,2\pi)=\order{1/g^0}\ .
\end{cases}
\]
Particles within different regimes can, in principle, 
scatter with themselves or with other types of particles, 
but it is expected that their scattering phase 
is suppressed by powers of the coupling constant.
The MT elementary excitations and HM giant magnons also serve as
constituents for Frolov-Tseytlin spinning strings \cite{Frolov:2002av,Frolov:2003qc}
and Gubser-Kle\-ba\-nov-Polyakov spinning strings \cite{Gubser:1998bc}, respectively.

At weak coupling we find two regions for particles with real momenta.
These correspond to magnons and magnon-holes:
\[
g\to 0\quad \Longrightarrow \quad
\mbox{particle} \in
\begin{cases}
\mbox{magnon} & \mbox{if } C>0\ ,
\\
\mbox{hole} & \mbox{if } C<0\ .
\end{cases}
\]

An additional complication is that both at strong and at weak coupling
some branch points of the phase,
for example the ones that will be discussed in section \ref{sec:oddproof},
move outside the kinematical regime and thus
become inaccessible. 
The associated monodromies will then lead to additional integer 
labels for the particles in a perturbative treatment.
For instance, at strong coupling the real period grows infinitely large
with respect to the imaginary one. In this way periodicity along the real axes
is lost and the momentum $p$ is confined to a specific region. 
It will turn out that a shift by $2\pi$
can change the dressing factor. Therefore we have to specify
for all particles what multiple of $2\pi$ we are considering
in order to pin down the phase.
Similarly, the two types of particles at weak coupling
are related by the antipode map. The double antipode map is non-trivial
and the dressing phase does change under it. Therefore we have to distinguish
between magnons and their images under the double antipode map.
As we shall see, both at strong and weak coupling, there will
be more discrete choices to be made which lead to additional
labels for the distinct regimes of particles.
For the sake of clarity we shall not write out these labels explicitly.

\section{Crossing-Symmetric Series}
\label{sec:series}

In this section we will search for a general solution to the crossing equation of the form
\[\label{eq:strongexpand}
\theta=\sum_{n=0}^\infty \theta^{(n)}+\theta\supup{hom} \, ,
\]
where the summands $\theta^{(n)}$ represent
a $n$-loop contribution in the perturbative string world sheet theory at strong
coupling.
We will postpone the discussion of the homogeneous piece to a later section
 and focus in what follows on one particular solution of
the crossing relation given by the $\theta^{(n)}$.

\subsection{Series Representation}
\label{sec:seriesrep}

A reasonably general form of the dressing factor $\sigma$
is \cite{Arutyunov:2004vx,Beisert:2005wv}
\[
\label{eq:phase}
\sigma_{12}=\exp i\theta_{12},\qquad
\theta_{12}=
\sum_{r=2}^\infty
\sum_{s=r+1}^\infty
c_{r,s}\bigbrk{q_r(\xpm{1})\,q_s(\xpm{2})-q_s(\xpm{1})\,q_r(\xpm{2})} \ ,
\]
where the magnon charges are defined as
\[\label{eq:magcharge}
q_r(\xpm{})=\frac{i}{r-1}\,\lrbrk{\frac{1}{(\xp{})^{r-1}}-\frac{1}{(\xm{})^{r-1}}} \ ,
\]
and $c_{r,s}$ are some real coefficients depending on the 't~Hooft
coupling constant.

Before we proceed, let us motivate why the above form of the phase is useful:
First of all, the phase is defined purely in terms of magnon charges
which form a natural basis of conserved quantities.
Secondly, zero-momentum particles representing symmetry generators have
a trivial dressing factor. In addition, the first derivative of the phase around zero
momentum vanishes. These two properties are required for the correct realisation of
$\alg{psu}(2,2|4)$ symmetry, see e.g.~\cite{Beisert:2005fw}.
Thirdly, the phase is naturally doubly periodic on the complex torus.
The form \eqref{eq:phase} thus represents a basis of periodic two-parameter
functions with a couple of desired additional symmetry properties.
It can also be viewed as a mode decomposition for functions on a torus;
it is somewhat similar to a Fourier decomposition but with two periods.%
\footnote{Let us note that, although the function is
periodic by definition, infinite sums may lead to branch cuts which may render the analytic
continuation of the function aperiodic. We shall be interested in this type of 
analytic continuation.} Fourth, an analysis of perturbative integrable spin chains gives
\eqref{eq:phase} as the most general expression \cite{Beisert:2005wv}.
Although this is not directly applicable
to string theory models, we consider it a valid indication.
Finally, this form collaborates nicely with the scattering of bound states
in the bootstrap approach \cite{Roiban:2006gs,Chen:2006gq}.

The form of the above phase \eqref{eq:phase,eq:magcharge}
suggests to write it as
a symmetrisation of a function $\chi(x_1,x_2)$ \cite{Arutyunov:2006iu}
\<\label{chi}
\theta_{12}\eq
+\chi(\xp{1},\xp{2})
-\chi(\xp{1},\xm{2})
-\chi(\xm{1},\xp{2})
+\chi(\xm{1},\xm{2})
\nl
-\chi(\xp{2},\xp{1})
+\chi(\xm{2},\xp{1})
+\chi(\xp{2},\xm{1})
-\chi(\xm{2},\xm{1})\ .
\>
We will generally use this definition of $\theta$ in terms of $\chi$.
In order to match with \eqref{eq:phase,eq:magcharge} we have to set
\[
\chi(x_1,x_2)=\sum_{r=2}^\infty\sum_{s=r+1}^\infty
\frac{-c_{r,s}}{(r-1)(s-1)}\,\frac{1}{x_1^{r-1}x_2^{s-1}}\ .
\label{chisum}
\]
%

\subsection{Strong-Coupling Expansion}

The coefficients $c_{r,s}$ in \eqref{eq:phase} are known to first order at
strong coupling \cite{Arutyunov:2004vx,Beisert:2005cw,Hernandez:2006tk,Freyhult:2006vr}. At leading order they are given by
\cite{Arutyunov:2004vx}
\[\label{eq:coeffleading}
c^{(0)}_{r,s}=g\,\delta_{r+1,s} \ .
\]
The first quantum correction turns out to be \cite{Hernandez:2006tk}
\[\label{eq:coeffsubleading}
c^{(1)}_{r,s}=\frac{(-1)^{r+s}-1}{\pi}\,\frac{(r-1)(s-1)}{(r+s-2)(s-r)}\, ,
\]
which follows from a one-loop comparison
with spinning string energies \cite{Beisert:2005cw,Hernandez:2006tk,Freyhult:2006vr}.

It was shown in \cite{Arutyunov:2006iu} that the first two contributions
can be summed up to analytic expressions.
The leading order contribution
is given by
\[\label{zero}
\chi^{(0)}(x_1,x_2)=-\frac{g}{x_2}+g\lrbrk{-x_1+\frac{1}{x_2}}\log\lrbrk{1-\frac{1}{x_1x_2}} \ ,
\]
and the first order reads
\<\label{odd}
\chi^{(1)}(x_1,x_2)\eq
-\frac{1}{2\pi}\Li_2\frac{\sqrt{x_1}-1/\sqrt{x_2}}{\sqrt{x_1}-\sqrt{x_2}}
-\frac{1}{2\pi}\Li_2\frac{\sqrt{x_1}+1/\sqrt{x_2}}{\sqrt{x_1}+\sqrt{x_2}}
\nl
+\frac{1}{2\pi}\Li_2\frac{\sqrt{x_1}+1/\sqrt{x_2}}{\sqrt{x_1}-\sqrt{x_2}}
+\frac{1}{2\pi}\Li_2\frac{\sqrt{x_1}-1/\sqrt{x_2}}{\sqrt{x_1}+\sqrt{x_2}}\ ,
\>
where $\Li_2(z)$ is the dilogarithm function.%
\footnote{We have made use of the
identity $\Li_2(z)+\Li_2(1-z)=\sfrac{1}{6}\pi^2-\log(z)\log(1-z)$
to absorb all the terms bilinear in logarithms that appear in \protect\cite{Arutyunov:2006iu}.}

These contributions have recently been shown to satisfy
the crossing relation up to order
$\order{1/g^3}$ for MT excitations \cite{Arutyunov:2006iu}.
Furthermore in \cite{Beisert:2006zy} it was argued that the $n=1$ contribution to the phase
is actually sufficient to satisfy exactly the odd crossing relation
\eqref{crossoddeven,hoddeven}
for finite values of the coupling.
We shall provide a full proof of that statement in section \ref{sec:oddproof}.
It was also demonstrated that in order
to satisfy the original crossing relation \eqref{cross}
further even-$n$ contributions are needed.
We may therefore identify
\[
\theta\supup{odd}=\theta^{(1)}
\quad\mbox{and}\quad
\theta\supup{even}=\sum_{n=0}^\infty\theta^{(2n)}.
\]
%

\subsection{Proposal}

The central result of this paper is a proposal for
the coefficients $c^{(n)}_{r,s}$ in \eqref{eq:phase}
with even $n\geq 2$ such that the crossing relation \eqref{cross} is
satisfied. These take the following form
\<
\label{coeffs}
c_{r,s}^{(n)}\eq
\frac{1}{g^{n-1}}\,\frac{\bigbrk{(-1)^{r+s}-1}\bernoulli_{n}}
{4\cos(\half\pi n)\,\gammafn[n+1]\,\gammafn[n-1]}\,
\nl\qquad
\times
(r-1) (s-1)\,
\frac{\gammafn[\half(s+r+n-3)]}{\gammafn[\half(s+r-n+1)]}\,
\frac{\gammafn[\half(s-r+n-1)]}{\gammafn[\half(s-r-n+3)]}\ ,
\>
where $\bernoulli_n$ denotes the $n$-th Bernoulli number.
Note that $c^{(n)}_{r,s}=0$ 
if $r+s$ is even or if $n\geq s-r+3$.
The factor $(r-1)(s-1)$ cancels the denominators
in $q_r$ and $q_s$, and the sum over $r$ and $s$ can be performed easily. 
For every $\theta^{(n)}$ we find some rational
function in $\xpm{1,2}$.
For the first two terms in the expansion of $\chi$ we obtain
the following rational functions
\<\label{chiexp}
\chi^{(2)}(x_1,x_2)\eq -\frac{x_2}{24g(x_1x_2-1)(x_2^2-1)}\,,\nln
\chi^{(4)}(x_1,x_2)\eq -\frac{x_2^3+4x_2^5-9x_1x_2^6+x_2^7+3x_1^2x_2^7-3x_1x_2^8+3x_1^2x_2^9}
{720g^3(x_1x_2-1)^3(x_2^2-1)^5} \ .
\>
We have obtained similar expressions up to $\chi^{(12)}$, but they
are too bulky to be presented here.

A few features of the coefficients are worth mentioning:
The fact that the odd Bernoulli numbers are zero
relates nicely to the fact that odd-$n$ contributions are not
required for a solution of crossing.%
\footnote{In fact this is not straightforward
because \protect\eqref{coeffs} contains $\cos(\half\pi n)$ in
the denominator, which is zero for odd $n$.
Thus the coefficients are ambiguous for odd $n$, and we may only define
them to be zero. We will return to this issue in section \protect\ref{sec:homo}.}
The notable exception is $n=1$ with $\bernoulli_{1}=-\half$.
Remarkably, the properly regularised expression for \eqref{coeffs}
with $n=1$ yields \emph{precisely} \eqref{eq:coeffsubleading}!
Also the leading order coefficients \eqref{eq:coeffleading}
are contained in \eqref{coeffs} as the regularised
contribution at $n=0$.%
\footnote{The contribution with $r=1$, $s=2$ is zero
for all $n$ with the exception of $n=0$ where it is
defined ambiguously. Adding this contribution
with $c^{(0)}_{1,2}=g$ to the sum \protect\eqref{chisum} 
solves $h\supup{even}$ without the only term which makes direct reference
to $\xpm{}$, cf.~$x_2^-/x_2^+$.}
Therefore \eqref{coeffs} can be considered a natural
extension of \eqref{eq:coeffleading,eq:coeffsubleading} to higher orders.
Finally we mention that each coefficient $c_{r,s}(g)$ has a finite expansion in $1/g$
with the last contribution at $n=s-r+1$.

\subsection{Proof of Odd Crossing}
\label{sec:oddproof}

We now turn towards confirming the crossing relation for
our proposed series.
In this section we will prove that the odd crossing relation
\eqref{crossoddeven,hoddeven} is satisfied by $\theta^{(1)}$ alone.
We will address the proof of this statement in two steps. First we
will show that $\theta^{(1)}$ satisfies double crossing \eqref{dcross}, and
afterwards we will turn to the odd crossing relation.

\paragraph{Double Crossing.}

Under double crossing the $\xpm{}$ variables
are mapped to themselves:
$\xpm{}\mapsto 1/\xpm{}\mapsto\xpm{}$.
We therefore have to investigate the monodromies of the phase.
We will discard shifts by multiples of $2\pi$
because they will drop out after exponentiating the phase.

The phase $\theta^{(1)}$ is composed from dilog functions, therefore
let us review its monodromies first.
It has the following two:
When $z$ is taken once around $z=1$ (counterclockwise)
the analytic continuation of $\Li_2(z)$ shifts by
\[\label{dilogmono}
\oint_{z=1} d\Li_2(z)=-2\pi i \log(z)\ .
\]
Likewise for circles around $z=\infty$ it shifts by the
same amount in the opposite direction,
\[
\oint_{z=\infty}d\Li_2(z)=+2\pi i \log(z)\ .
\]
Altogether the sum of shifts for all points cancels as it should.

Equipped with these formulae we can now consider the monodromies of $\chi^{(1)}$.
The relevant points are those where the argument of the dilog
is $1$ or $\infty$.
This happens at $x_1=\infty$, $x_2=\pm 1,0$ or $x_1=x_2$.
The monodromies at $x_2=\pm 1$ are
\[\label{monochi1}
\oint_{x_2=\pm 1}d\chi^{(1)}(x_1,x_2)=\pm i\log\frac{x_1-1/x_2}{x_1-x_2}\ .
\]
More explicitly this means that the monodromies at $\sqrt{x_2}=+1$ and
$\sqrt{x_2}=-1$ both take the above value
with the $+$ sign. Similarly for $\sqrt{x_2}=\pm i$ and the $-$ sign.
The monodromy for $x_1=x_2$ however needs to be split
into the two cases $\sqrt{x_2}=+\sqrt{x_1}$ and
$\sqrt{x_2}=-\sqrt{x_1}$ for which we get opposite monodromies
\[\label{monochi2}
\oint_{\sqrt{x_2}=\pm\sqrt{x_1}}d \chi^{(1)}(x_1,x_2)=
\pm i\log\frac{1+1/\sqrt{x_1}\sqrt{x_2}}{1-1/\sqrt{x_1}\sqrt{x_2}}\ .
\]
The potential monodromies at $x_1=\infty$ and $x_2=0$ both cancel out.

We are finally in the position to consider double crossing
of $\theta^{(1)}_{12}$. The monodromy \eqref{monochi2}
cannot contribute here because it is symmetric under
the interchange of $x_1$ and $x_2$ whereas
$\theta^{(1)}_{12}$ is anti-symmetric.
In other words, the monodromies
for $\sqrt{x_1}$ circling around $\pm\sqrt{x_2}$
cancel out between each term $\chi^{(1)}(x_1,x_2)$
and the corresponding term $-\chi^{(1)}(x_2,x_1)$ in \eqref{chi}.
We are thus left with \eqref{monochi1}.
The monodromies of $\theta_{12}$ for $\xp{1}=\pm 1$
and $\xm{1}=\pm 1$ are
\<\label{thetamono}
\oint_{\xp{1}=\pm 1}
d\theta^{(1)}_{12}\eq
\mp i\log\frac{\xm{2}-\xp{1}}{\xp{2}-\xp{1}}\,
         \frac{\xp{2}-1/\xp{1}}{\xm{2}-1/\xp{1}}\ ,
\nln
\oint_{\xm{1}=\pm 1}
d\theta^{(1)}_{12}\eq
\mp i\log\frac{\xp{2}-\xm{1}}{\xm{2}-\xm{1}}\,
         \frac{\xm{2}-1/\xm{1}}{\xp{2}-1/\xm{1}}\ .
\>

Now we need to investigate what path $\xpm{1}$
takes under the double crossing map.
Let us parametrise the momentum $p$ using
Jacobi's amplitude function $\ellAM$ with elliptic modulus $k$
and rapidity variable $z$,
\[\label{eq:momtorus}
p=2\ellAM(z,k),\qquad k=4ig \ .
\]
Then the half-periods of the torus are given by
\[
\omega_1=2\ellK(k),\qquad
\omega_2=2i\ellK(\sqrt{1-k^2})-2\ellK(k) \ .
\]
We note that $p$ increases by $4\pi$ under a shift
of the full real period $2\omega_1$. Therefore this parametrisation of
the torus is a double covering.
A single covering can be achieved as well, but at the cost of
expressions which are substantially longer than e.g.~\eqref{eq:momtorus}.

Double crossing takes the rapidity $z$ once around the imaginary period of the torus.
The direction is not immediately obvious, it could be either of the
two maps $z\mapsto z\pm 2\omega_2$. Let us for convenience assume the positive sign.
To define a path between the two points
we shall furthermore assume that the real part of $z$ remains constant.
Then it turns out that
for $\Re z\in (-\frac{1}{4}\omega_1,+\frac{1}{4}\omega_1)+\omega_1\Integers$
the variables $\xpm{}$ circle clockwise around $+1$.
Conversely, for $\Re z\in (\frac{1}{4}\omega_1,\frac{3}{4}\omega_1)+\omega_1\Integers$
they circle counterclockwise around $+1$.
Here we take $-1$ as the reference point for defining
the outside of the path.

The overall monodromy is therefore
\[
\theta^{(1)}_{\bar{\bar1}2}-\theta^{(1)}_{12}=\pm i\log (h\supup{odd}_{12})^2.
\]
The plus sign is valid for $\Re z\in (-\frac{1}{4}\omega_1,+\frac{1}{4}\omega_1)+\omega_1\Integers$
(roughly speaking for $p\approx 2\pi\Integers$)
and agrees literally with \eqref{dcross}.
The minus sign holds for $\Re z\in (\frac{1}{4}\omega_1,\frac{3}{4}\omega_1)+\omega_1\Integers$
(roughly speaking for $p\approx \pi+2\pi\Integers$)
and we should therefore reverse the crossing path for these momenta
in order to achieve agreement with \eqref{dcross}.
This is not a problem as the definition of the
crossing relation in fact allows both signs in
$z\mapsto z\pm \omega_2$. Nevertheless, it would be illuminating
to find out why we have to choose different orientations
depending on the particle momentum.
Alternatively, we could specify a path such that $\xpm{}$ always circle
clockwise around $+1$.
This completes the proof that $\theta^{(1)}$ satisfies the double crossing
relation \eqref{dcross}.

\paragraph{Odd Crossing.}

Above we have investigated the structure of monodromies
of the function $\chi^{(1)}(x_1,x_2)$.
This was sufficient for the proof of double crossing
because the double antipode map takes $x$ to itself.
The regular crossing relation on the other hand maps
$x$ non-trivially and therefore monodromies are not
sufficient for proving the full relation. Nevertheless
they are essential for our understanding:

We have seen for instance that
there are monodromies at $x_k=\pm 1$
and that the monodromies at
$\sqrt{x_1}=\pm\sqrt{x_2}$ cancel in the combination
$\chi^{(1)}(x_1,x_2)-\chi^{(1)}(x_2,x_1)$.
Thus, the symmetrised combination has a simpler analytic structure
and one should be able to simplify the expression \eqref{odd}.
We obtain the following form
\[\label{oddnew}
\chi(x_1,x_2)-\chi(x_2,x_1)
=\psi(q_1-q_2)+\frac{\Li_2(x_2)-\Li_2(-x_2)-\Li_2(x_1)+\Li_2(-x_1)}{2\pi}\ ,
\]
with the auxiliary function $\psi(q)$
\[\label{oddaux}
\psi(q)=\frac{1}{2\pi}\Li_2\bigbrk{1-e^{iq}}
-\frac{1}{2\pi}\Li_2\bigbrk{1-e^{iq+i\pi}}
-\frac{i}{2}\log\bigbrk{1-e^{iq+i\pi}}
+\frac{\pi}{8}\ ,
\]
and where $q_k$ is related to $x_k$ as follows
\[\label{qvar}
e^{iq}=\frac{x+1}{x-1}\ .
\]
Note that all the terms besides $\psi(q_1-q_2)$ in
\eqref{oddnew} depend on either $x_1$ or $x_2$ only
and thus they cancel out in the dressing phase \eqref{chi}
\[
\theta^{(1)}_{12}=
\psi(q^+_1-q^+_2)
-\psi(q^+_1-q^-_2)
-\psi(q^-_1-q^+_2)
+\psi(q^-_1-q^-_2)\,.
\]
Here $q^\pm$ are related to $\xpm{}$ as in \eqref{qvar}.

A very tedious method to compare \eqref{oddnew}
to \eqref{odd} is to apply dilogarithm identities
such as the Abel identity.%
\footnote{We have used 16 Abel identities
to show the equivalence, but for a better choice of identities
there may be a shorter path.}
However, it is much easier to confirm that \eqref{oddnew}
has the right monodromies. Here the variable $q$ comes into play.
It shifts by $+2\pi$ for a full clockwise rotation around $x=+1$ w.r.t.~$x=-1$.
This matches nicely with the monodromies of the dilogs in $\psi(q)$,
cf.~\eqref{dilogmono},
namely
\[\label{dilogshift}
\Li_2\bigbrk{1-e^{iq+2\pi i n}}=\Li_2\bigbrk{1-e^{iq}}-2\pi i n\log \bigbrk{1-e^{iq}} \ .
\]
In this new notation, the double crossing map reads
\[\label{doublenew}
\psi(q+2\pi)
=\psi(q)-i \log \frac{1-e^{iq}}{1+e^{iq}}
=\psi(q)-i \log \bigbrk{-i\tan(\half q)}\ ,
\]
where
\[
i\tan(\half q_1-\half q_2)=\frac{x_1-x_2}{1-x_1x_2}\ .
\]
After multiplying the various terms for $\xpm{1,2}$ in
\eqref{chi} this function becomes
\[
\frac{\xp{1}-\xp{2}}{1-\xp{1}\xp{2}}\,
\frac{1-\xp{1}\xm{2}}{\xp{1}-\xm{2}}\,
\frac{1-\xm{1}\xp{2}}{\xm{1}-\xp{2}}\,
\frac{\xm{1}-\xm{2}}{1-\xm{1}\xm{2}}
=\frac{1}{(h\supup{odd}_{12})^{2}} \ ,
\]
in agreement with the double crossing relation \eqref{crossoddeven,hoddeven}.

Before we turn towards the odd crossing relation, let us investigate
unitarity and parity invariance of the above expression \eqref{oddnew}.
Both translate to $\psi(q)$ being an odd function in $q$ which is not manifest.
To prove it, we have to use the dilog identity
\[
\Li_2(1-e^{i q})+\Li_2(1-e^{-iq})=\half q^2
\]
to flip the sign of the exponents in the dilogs.
Before we can do this in the second term in \eqref{oddaux}
we have to shift via \eqref{dilogshift}.
The remainder of the proof reads as follows
\<
\psi(q)+\psi(-q)\eq
+\frac{q^2}{4\pi}
-\frac{(q+\pi)^2}{4\pi}
-\frac{i}{2}\log\bigbrk{1+e^{iq}}
+\frac{i}{2}\log\bigbrk{1+e^{-iq}}
+\frac{\pi}{4}
\nln\eq
+\frac{q^2}{4\pi}
-\frac{(q+\pi)^2}{4\pi}
+\frac{q}{2}
+\frac{\pi}{4}=0\ .
\>

Finally, we can attack the odd crossing relation.
The discussion at the end of the double crossing proof
leads to the conclusion that the variable $q$
as defined in \eqref{qvar} shifts by $+2\pi$
under double crossing.
The sign for shifts of $q$ has to be positive in all cases.
This is in contradistinction to the shift
in the above rapidity variable $z$
for which the path on the universal cover of the
torus has to be chosen carefully. Thus $q$ appears to be a more
fundamental quantity than $z$.
Under regular crossing $e^{iq}$ maps to $-e^{iq}$
and thus $q$ has to shift by $+\pi$ to match with the above.
Let us see how $\psi(q)$ behaves under such a shift.
After cancelling an intermediate
term and shifting according to \eqref{dilogshift} we find 
\<
\psi(q)+\psi(q+\pi)
\eq
\frac{i}{2}\log\bigbrk{1-e^{iq}}
-\frac{i}{2}\log\bigbrk{1+e^{iq}}
+\frac{\pi}{4}
\nln\eq
-\frac{i}{2}\log\bigbrk{i\cot(\half q)}
+\frac{\pi}{4}\,.
\>
In analogy to \eqref{doublenew}
this proves the odd crossing relation.

As the odd part of the crossing relation is solved,
we can focus on the even part for the remainder of this paper.

\subsection{Confirmation of Even Crossing}

We will now provide evidence that our proposed coefficients \eqref{coeffs}
satisfy also the even crossing symmetry. 
For the $n$-loop expressions, $n$ even,
we find the following contributions to the crossing relation
\<\label{crossphase}
\theta^{(0)}_{12}+\theta^{(0)}_{\bar 12}\eq
i\log\frac{\xp{2}}{\xm{2}}
+g\Delta\log\frac{\Delta^2+1/g^2}{\Delta^2}+i\log\frac{\Delta+i/g}{\Delta-i/g}\ ,
\nln
\theta^{(n)}_{12}+\theta^{(n)}_{\bar 12}\eq
-\frac{i^{n}\bernoulli_{n}}{n(n-1)\,g^{n-1}}
\lrbrk{\frac{2}{\Delta^{n-1}}-\frac{1}{(\Delta+i/g)^{n-1}}-\frac{1}
{(\Delta-i/g)^{n-1}}} \, ,
\>
with $\Delta=u_1-u_2$.
These expressions are exact, therefore they apply
in any of the strong-coupling regimes, but we have made
use of the defining identity of $\xpm{}$ \eqref{def}.
Although we do not have a general proof of the last
expression in \eqref{crossphase}, we have confirmed it up to $n=12$.
The even crossing phase to compare to reads
\[\label{even}
-i\log h\supup{even}_{12}
=i\log\frac{\xp{2}}{\xm{2}}+\frac{i}{2}\log\frac{\Delta+i/g}{\Delta-i/g} \, .
\]
Our claim is that
\[\label{crossclaim}
-i\log h\supup{even}
=
\sum_{n=0}^N
\bigbrk{
\theta^{(2n)}_{12}
+\theta^{(2n)}_{\bar 12}
}
+\order{1/g^{2N+3}} \, ,
\]
for any upper limit of the sum $N$.
Assuming the validity of \eqref{crossphase}
it is easy to verify \eqref{crossclaim} to very large values of $N$.
However, the sum in \eqref{crossclaim} is in fact problematic if we set $N=\infty$
to confirm the exact crossing relation.
In that case we cannot, a priori, interchange the expansion in
$1/g$ and the summation. This is related to the fact that
\eqref{crossphase,crossclaim} is not a pure power series.
In fact, it is straightforward to
show that the sum does not converge for arbitrarily large values of $g$.
The reason is that the Bernoulli numbers $\bernoulli_n$
grow like $n!$ and thus faster than $a^n$ for any number $a$.
An alternative path for testing crossing symmetry 
is to Borel sum the above series.
Remarkably, this leads to precise agreement with \eqref{even}.
We believe that this is convincing evidence for the validity
of our proposed solution.

It is conceivable that the lack of convergence of \eqref{crossphase}
also implies that the original series defining the phase
\eqref{chi,chisum,coeffs} is problematic. 
In order to investigate the phase, in particular its analytic structure,
we would therefore benefit very much from a more explicit representation
of the sum. This will be the topic of the next section.
In \appref{sec:WeakBorel} we will present another representation of
the series which is most useful for weak coupling.

\section{Crossing-Symmetric Function}
\label{sec:function}

In this section we will present an analytic expression for the resummed series
which allows us to study the structure of singularities in the phase.
In particular, we find exact expressions for the bound state poles 
of giant magnons recently derived in \cite{Hofman:2006xt}.

\subsection{Claim}

Our claim is that the proper analytic
expression corresponding to the above series is
\<\label{chianalytic}
\chi\supup{even}(x_1,x_2)\eq
\lim_{N\to\infty}\left[
\frac{g}{2x_1}\log \frac{gx_2}{N}
-\frac{i}{4}\sum^N_{n=1}
\log \frac{1-1/x^{}_1\xexp{+2n}{}(x_2)}{1-1/x^{}_1\xexp{-2n}{}(x_2)}\right]
\nlnum\nonumber
+\frac{g}{2}\lrbrk{-\frac{1}{x_1}-\frac{1}{x_2}}
+\frac{g}{2}\lrbrk{-x_1-x_2+\frac{1}{x_1}+\frac{1}{x_2}}\log\lrbrk{1-\frac{1}{x_1x_2}}\ .
\>
Here $N$ is a cut-off parameter
which should be taken to infinity.
In that limit, the first term correctly regularises the logarithmically divergent sum.
The terms on the second line are such that they give no
contribution to the physical phase $\theta\supup{even}$ via \eqref{chi},
but they are necessary for reproducing correctly the expected behaviour of the function $\chi$
in the series representation
given in the previous section. 
Note also that the explicit appearance of the cut-off $N$
in the first term is of this sort.
The divergence of the sum would thus cancel out
the full phase $\theta\supup{even}$ even without introducing a regularisation.
The analytic function $\chi\supup{even}$
is expressed in terms of the new
quantities $\xexp{n}{}$, which are related to $x$ as
\[\label{xndef}
\xexp{n}{}+\frac{1}{\xexp{n}{}}-x-\frac{1}{x}=\frac{in}{2g} \, .
\]
%

\subsection{Gluing Two Infinite Genus Surfaces}

Before we compare the function $\chi\supup{even}$ to the series
expression of the previous section,
we would like to investigate its
analytic structure.
First, consider the map $\Comp\to\Comp^\infty$
\[\label{xxnmap}
x\mapsto \bigbrk{\ldots,\xexp{-2}{},\xexp{-1}{},\xexp{0}{},\xexp{+1}{},\xexp{+2}{},\ldots}\ .
\]
This map has branch points where $x+1/x+in/2g=\pm 2$ for any integer $n$.
When $x$ moves once around one of these branch points,
the component $\xexp{n}{}$ is mapped to its inverse,
which is the other solution of \eqref{xndef}.
All the other components will remain unchanged.
The only exception is the map $\xexp{0}{}(x)$ which we shall
define as the identity map
\[\label{x0}
\xexp{0}{}(x):=x \ .
\]
This is possible because the branch points degenerate for $n=0$.
The analytic completion of the map thus has infinitely many branch points
with distinct monodromies
and is defined on a Riemann surface of infinite genus.
The function $\chi\supup{even}$ is essentially defined with $x_2$ on
this Riemann surface because for every even $n$ we have
made a choice between $\xexp{n}{}$ and its inverse.
The fact that we only require even $n$ does not alter the
picture qualitatively, therefore let us stick to the more general surface.

Next we need to consider $\theta\supup{even}_{12}$ which consists of
terms of the sort $f(\xp{})-f(\xm{})$, cf.~\eqref{chi}.
The problem is now that these functions are defined on two different
Riemann surfaces. 
In general, the combined function would
be defined on the product of two infinite-genus surfaces.
Nevertheless, the two variables $\xp{}$ and $\xm{}$
are related in a special way such that all the branch points of
$f(\xp{})$ and $f(\xm{})$ coincide.
Therefore the mono\-dromies of the two functions are not unrelated.
Namely, if we move $\xp{}$ once around a branch point that
inverts $\xexp{n-1}{}(\xp{})$ then $\xm{}$ has to move
around the corresponding branch point that inverts $\xexp{n+1}{}(\xm{})$.
Consequently, it is consistent to make a specific choice
\[\label{matchmap}
\xexp{n-1}{}(\xp{})=\bigbrk{\xexp{n+1}{}(\xm{})}^{s^{(n)}}\ ,
\]
where $s^{(n)}=\pm 1$ determines whether the $\xexp{n}{}$'s
are inversely related or not.%
\footnote{Notice that \protect\eqref{matchmap} for $n=1$ is not in conflict 
with \protect\eqref{x0}. Since $x^+$ and $x^-$ are related by \protect\eqref{def}, 
moving $x^-$ around the branch cut of $x^{(2)}(x^-)$ will also invert $x^+$. 
This does not need to affect $\xexp{n}{}(x^+)$ for $n\neq 0$, because 
they depend of the combination $x+1/x$.}
Moreover we are forced to make this choice in order to define the function
$f(\xp{})-f(\xm{})$ properly because variations of $\xpm{}$
cannot flip the signs $s^{(n)}$.
In conclusion, the combination $f(\xp{})-f(\xm{})$
requires us to fix the signs $s^{(n)}$ and is
then defined on an infinite-genus surface.
In contrast, the function $f(x)-f(y)$ with uncorrelated
$x,y$ is uniquely defined on the product of two
infinite-genus surfaces without sign ambiguities.

It is important to stress that we should consider
not only the $\xpm{}$'s but also the signs $s^{(n)}$, at least for
$n$ odd, as kinematic parameters of the particle.
Particles with different signs are not equivalent to each other,
which manifests in a different scattering behaviour.
For simplicity of notation, using \eqref{matchmap}, we can systematically write
$\xexp{n-1}{}(\xp{k})$ in terms of $\xexp{n+1}{}(\xm{k})$.
Consequently we shall define a single set of kinematic parameters
$\xexp{n}{k}$
which are related to $\xexp{n}{}(\xpm{k})$ by
\[
\xexp{n}{}(\xm{k})=\xexp{n-1}{k},\qquad
\xexp{n}{}(\xp{k})=(\xexp{n+1}{k})^{s^{(n+1)}_k}.
\]
with $\xexp{-1}{k}=\xm{k}$, but not necessarily
$\xexp{+1}{k}=\xp{k}$.

We should consider the transformation of the new parameters under
the discrete symmetries of the system,
in particular parity and the antipode map, cf.~\eqref{paritymap,crossmap}.
The antipode simply maps the $\xexp{n}{k}$ to their inverse
\[
\xexp{n}{k}\mapsto 1/\xexp{n}{k}.
\]
Parity maps $\xpm{}\mapsto-\xmp{}$
and it is consistent to define
$\xexp{n}{}(-x)=-\xexp{-n}{}(x)$.
Hence parity will act on the
$\xexp{n}{k}$ and on the signs $s^{(n)}$
as
\[\label{parity}
\xexp{n}{k}\mapsto -(\xexp{-n}{k})^{s^{(n)}_k} \, ,\qquad
s^{(n)}\mapsto s^{(-n)} \, ,
\]
implying that particles with $s^{(+n)}\neq s^{(-n)}$ do not map to
themselves under parity.

After making the choice of signs $s^{(n)}_{1,2}$ for each particle
we can compute the phase $\theta\supup{even}_{12}$
\<\label{fulleven}
\theta\supup{even}_{12}
\eq
-\frac{i}{2}\log \frac{1-1/\xm{1}\xp{2}}{1-1/\xp{1}\xm{2}}
+\frac{g}{2}\lrbrk{\frac{1}{\xp{1}}-\frac{1}{\xm{1}}}\log \frac{\xp{2}}{\xm{2}}
-\frac{g}{2}\lrbrk{\frac{1}{\xp{2}}-\frac{1}{\xm{2}}}\log \frac{\xp{1}}{\xm{1}}
\nl
-\frac{i}{4}\sum^\infty_{n=-\infty} \sign(2n-1) \,
\frac{1-s^{(2n-1)}_1}{2}
\log
\frac{\xm{2}-\xexp{2n-1}{1}}{\xp{2}-\xexp{2n-1}{1}}\,
\frac{\xp{2}-1/\xexp{2n-1}{1}}{\xm{2}-1/\xexp{2n-1}{1}}
\nlnum
-\frac{i}{4}\sum^\infty_{n=-\infty}\sign(2n-1) \,
\frac{1-s^{(2n-1)}_2}{2}
\log
\frac{\xp{1}-\xexp{2n-1}{2}}{\xm{1}-\xexp{2n-1}{2}}\,
\frac{\xm{1}-1/\xexp{2n-1}{2}}{\xp{1}-1/\xexp{2n-1}{2}}  \, , \nonumber
\>
This phase is consistent with unitarity 
if we notice that when exchanging the particles we should interchange the momenta
as well as the signs. It has also the right behaviour under parity.
Parity invariance implies that the overall scattering phase should change sign
under \eqref{parity}. The phase \eqref{fulleven} indeed fulfils this property.
We can now consider the crossing relation. 
It is straightforward to verify the even crossing relation
\eqref{crossoddeven,hoddeven}. In fact all the terms on the second and 
third line in \eqref{fulleven} represent homogeneous solutions of crossing.

For generic values of the signs, the phase \eqref{fulleven}
gives rise to square root singularities in the scattering
matrix. 
We cannot offer an explanation for this puzzling behaviour,
however, there are two points to be remarked: 
On the one hand one might consider giving up manifest parity invariance
and use the expression \eqref{eq:evenleft} for $\chi\supup{even}$ 
instead of \eqref{chianalytic} in which case there would be no 
fractional singularities.
On the other hand, one could adjust the signs $s^{(n)}$ 
such that the square root singularities go away.
This happens for scattering of parity self-conjugate particles
with equal sign assignations $s^{(n)}_1=s^{(n)}_2$, 
as well as for particles with all signs equal, $s^{(n)}_1=s_1$, $s^{(n)}_2=s_2$.
In these cases several terms appearing in \eqref{fulleven}
can be seen to cancel among themselves using
\<
\bigbrk{\xpm{k} -\xexp{n}{j}}
\bigbrk{1-1/\xpm{k}\xexp{n}{j}}
= u_k-u_j-\frac{i(n\mp1)}{2g} \, ,
\>
and we are left with a reduced expression whose singularities
lead to poles and zeros only. 
In particular a model 
whose particles have $s^{(n)}_k=s_k=\pm 1$ would be free
from fractional singularities, which also leads to 
simple expressions for the dressing phase.
We will argue in the next section that
the two choices $s_k=\pm 1$ are related 
to MT excitations and HM giant magnons respectively.

When all $s^{(n)}=+1$ the total phase depends on $\xpm{1}$ and $\xpm{2}$
only
\[\label{eq:evenanalytic}
\theta\supup{elem}_{12} = -\frac{i}{2} \log \frac{1-1/\xm{1}\xp{2}}{1-1/\xp{1}\xm{2}}
+\frac{g}{2}
\lrbrk{\frac{1}{\xp{1}}-\frac{1}{\xm{1}}}\log\frac{\xp{2}}{\xm{2}}
-\frac{g}{2}
\lrbrk{\frac{1}{\xp{2}}-\frac{1}{\xm{2}}}\log\frac{\xp{1}}{\xm{1}}\,,
\]
and is thus naturally defined on a torus. 
This choice produces a minimal
set of poles and zeros in the scattering matrix. 
We will therefore refer
to the phase above as ``elementary''.
It differs from the solution of the crossing relation proposed in
\cite{Beisert:2006zy} by a term $\delta\theta_{12}=\half(C_1 p_2-C_2 p_1)$.
This piece $\delta\theta_{12}$
repairs some of the manifestly unphysical behaviour
of the phase proposed in \protect\cite{Beisert:2006zy}.
Since $(C,p)\mapsto -(C,p)$ under the antipode map, it is clear that
$\delta\theta_{12}$ is a homogeneous solution of the crossing relation.
Remarkably, the phase \eqref{eq:evenanalytic} has appeared previously
in the context of light-cone gauge quantisation at leading order
in \cite{Frolov:2006cc}.%
\footnote{We thank M.~Staudacher as well
as S.~Frolov, M.~Zamaklar for pointing this out to us.}
As emphasised in this article, the rightmost term
in \eqref{eq:evenanalytic} combines nicely
with the phase contribution $-p_1L$ from the Bethe equations.
Together, the terms multiplying $p_1$ 
form the light-cone momentum $p_{+,2}$.
Although the term manifestly removes periodicity 
of the phase by shifts of momenta by $2\pi$, 
its appearance is actually useful because the length $L$ is not 
a physical quantity, whereas the light-cone momentum
is a charge under one of the Cartan generators
of $\alg{psu}(2,2|4)$.

The choice $s^{(n)}=-1$ leads to the following result
\[\label{fullevenHM}
\theta\supup{giant}_{12} =  \theta\supup{elem}_{12}
-\frac{i}{2}\sum^\infty_{n=-\infty} 
\sign(2n-1) \, \log
\frac{\xp{1}-\xexp{2n-1}{2}}{\xm{1}-\xexp{2n-1}{2}} \,
\frac{\xm{2}-\xexp{2n-1}{1}}{\xp{2}-\xexp{2n-1}{1}} \, .
\]
It gives rise to an infinite array of additional poles and zeros
in the scattering matrix. 
The proper definition of \eqref{fullevenHM}
needs the full-fledged, infinite-genus
surface introduced above.
We will connect it to the HM giant magnons and
thus denote it by ``giant''.

Before we make this more explicit by considering 
the strong coupling limit of the phase,
we would like to make a final comment regarding the choice of
signs $s^{(n)}$ for the Bethe equations: 
Although all assignments of $s^{(n)}$ 
may be meaningful to distinguish different kinds
of particles propagating on the infinite line, 
there should be one specific choice to be used 
for the Bethe equations in \cite{Beisert:2005fw}.
This is analogous to the situation for bound states of magnons
which can exist as elementary objects on the infinite line,
but not on the circle. 
We consider it very likely that 
one of the above two choices \eqref{eq:evenanalytic,fullevenHM}
is the correct one. 
On the one hand, the ``elementary'' choice has minimal genus
and would therefore lead to a relatively simple analytic structure
of the Bethe equations. 
On the other hand, the ``giant'' choice naturally incorporates
the giant magnon solitons and might be favourable
from a physics standpoint. 
However, as already contemplated in \cite{Hofman:2006xt},
the giant magnon may turn out to be a composite object;
below we will find some further indications strengthening this
point of view. In that case, the elementary choice $s^{(n)}=+1$ 
would most likely be the correct one for the Bethe equations.

\subsection{Strong Coupling}

In this section we will compare
the analytic expression \eqref{chianalytic} for $\chi\supup{even}$
to the perturbation series \eqref{chisum}.
The latter is defined at strong coupling where
the definition of the former simplifies drastically.

At strong coupling the above infinite-genus surface degenerates
into many disjoint regions.
This is related to the fact that
either $\xexp{n}{}=x+\order{1/g}$ or $\xexp{n}{}=1/x+\order{1/g}$.
Thus the function $\chi\supup{even}(x_1,x_2)$ does not have
infinitely many branch cuts anymore.
In other words, the branch points 
of the maps $x\mapsto \xexp{n}{}(x)$
have all moved to $1+\order{1/\sqrt{g}}$ where they
cannot be used to change sheets individually.%
\footnote{In fact there is a tail
of branch points for $n\sim g$ 
which extends throughout the complex plane.
This allows to invert contributions on a global scale
for very large $n$, but not for finite $n$.}
Therefore for every $n\neq 0$ we can definitely choose between
$\xexp{n}{}\approx x$ or $\xexp{n}{}\approx 1/x$.

Although the analytic function \eqref{chianalytic} intrinsically leads
to the previous infinite set of strong coupling choices, only
one of them directly connects with the perturbative series
\eqref{chisum,coeffs} for $\chi\supup{even}$. This is the simplest case
$\xexp{n}{}\approx x$. A restriction like this might seem unnatural,
however recall
that the perturbative sum does not 
converge literally. 
The interpretation
we are advocating for in this section is that once we associate an
analytic expression to the problematic sum, we unavoidably get
an enlarged phase space.

The agreement between the case $\xexp{n}{}\approx x$ and the series
\eqref{chisum,coeffs} can be checked using the Euler-MacLaurin
summation formula
\[
\sum_{n=1}^\infty f(n/g) = g \int_0^\infty dz\,f(z)
-
\sum_{k=1}^\infty \frac{\bernoulli_{2 k}}{(2 k)!\, g^{2k-1}}\, f^{(2k-1)}(0) \, ,
\label{EM}
\]
where $f^{(k)}$ denotes the $k$-th derivative of $f(z)$ and we have assumed
that the function vanishes at zero and infinity and in addition all its
derivatives vanish at infinity. In our case $f(z)$ is defined via%
\footnote{Note that $g$ does not 
appear explicitly in the function, cf.~\protect\eqref{xndef}.}
\[\label{function}
f(n/g) = -\frac{i}{4} \log
\frac{1-1/x^{}_1\xexp{+2n}{}(x_2)}{1-1/x^{}_1\xexp{-2n}{}(x_2)}
\, .
\]
The integral by itself is divergent.
The divergence is however cured by 
the counterterm in \eqref{chianalytic}
and we should compute the following finite combination
\<
\earel{}
\lim_{N\to\infty}
\left[
\frac{g}{2x_1}\log \frac{x_2}{N/g}+
g\int_0^{N/g} dz\,f(z)
\right]
\nln\eq
\frac{g}{2}\lrbrk{\frac{1}{x_1}-\frac{1}{x_2}}
+
\frac{g}{2}\lrbrk{-x_1+x_2-\frac{1}{x_1}+\frac{1}{x_2}}
\log\lrbrk{1-\frac{1}{x_1x_2}}.
\>
Together with the terms on the 
second line in \eqref{chianalytic} we reproduce exactly
$\chi^{(0)}$ of the AFS phase in \eqref{zero}.
Finally each term in the sum over derivatives
produces precisely $\chi^{(2k)}$, cf.~\eqref{chiexp},
which we have confirmed up to $\chi^{(12)}$.
Note that the presence of the Bernoulli numbers in the Euler-MacLaurin formula
nicely fits their appearance in our proposed coefficients.

\paragraph{MT Regime.}

As we have chosen $x^{(n)}\approx x$, 
which applies to $x=\xp{}$ as well as
$x=\xm{}$, all the $s^{(n)}$ in \eqref{matchmap} will coincide
being equal to either $+1$ or $-1$.
The MT excitations are 
characterised by $\xp{} \approx \xm{}$, see \eqref{MT},
and therefore we require all $s^{(n)}=+1$.
The scattering phase then reduces to the simple analytic expression \eqref{eq:evenanalytic}.
This phase agrees with all the available data for spinning
strings and near plane wave states,
i.e.~so far only the leading order in $\theta\supup{elem}$ 
\cite{McLoughlin:2004dh,Frolov:2006cc,Beisert:2005mq,Schafer-Nameki:2005tn}
together with the odd contribution $\theta\supup{(1)}$
\cite{Hernandez:2006tk,Freyhult:2006vr}.

\paragraph{HM Giant Magnons.}

The giant magnons have $\xp{} \approx 1/\xm{}$, see \eqref{HM}, implying that
this case should be represented by $s^{(n)}=-1$.
The cancellations that took place for MT excitations do not occur now and
we are left with an infinite array of poles and zeros in the scattering
matrix, as can be seen in \eqref{fullevenHM}.

Based on the connection to the sine-Gordon model \cite{Pohlmeyer:1975nb,Mikhailov:2005qv,Mikhailov:2005zd}
it was shown in \cite{Hofman:2006xt} that the semiclassical behaviour of the giant
magnons was described by the AFS dressing phase \cite{Arutyunov:2004vx}.
The latter reads in this limit%
\footnote{The sign conventions for the phase in \protect\cite{Hofman:2006xt} seem to be reversed from ours.}
\<\label{eq:HM0}
\theta_{12}^{(0)}\eq
2g\bigbrk{\cos(\half p_1)-\cos(\half p_2)}
\log\frac{\sin^2\bigbrk{\quarter(p_1-p_2)}}{\sin^2\bigbrk{\quarter(p_1+p_2)}}
+\order{1/g^0} \ .
\>
The leading phase \eqref{eq:HM0} has branch cuts starting from $p_1=\pm p_2$.
It is natural to interpret these branch cuts as condensates of poles and
zeros.%
\footnote{We thank N.~Dorey and J.~Maldacena for discussions and explanations.}
Notice that the square in the argument of the log implies that it has
two branch cuts originating from $p_1=p_2$, and correspondingly from
$p_1=-p_2$.
One of them is associated with poles and the other with zeros
depending on the sign of the prefactor of the log.

The poles of the scattering matrix have the interpretation of bound states.
The bound states corresponding to the phase \eqref{eq:HM0} were derived
in \cite{Hofman:2006xt}.
Recalling that $u=2\cos(\half p)$, they appear at 
\[\label{boundHM}
u_1-u_2=\frac{i n}{2 g} +\order{1/g^2}\ .
\]
Indeed, the semiclassical counting of these states
agrees with the discontinuity of the cut in \eqref{eq:HM0}.
Namely, the prefactor of the $\log$ increases by $i/2$
between the positions of 
two adjacent bound states.
We have shown above that the leading piece of $\chi\supup{even}$,
the integral term in the 
Euler-MacLaurin formula, leads to the AFS phase.
Therefore the array of poles and zeros in \eqref{fullevenHM} precisely reconstructs
the branch cuts in \eqref{eq:HM0}:
The exact scattering phase leads to double poles at
\[\label{boundus}
u_1-u_2= \frac{in}{g} \, .
\]
This is not in contradiction with the results of \cite{Hofman:2006xt}.
Relation \eqref{boundHM} was derived using semiclassical quantisation,
and can only be trusted for the overall counting of states,
but not for their precise positions. 
The result \eqref{boundHM} should be understood as 
an average density of approximately $2g$ poles per imaginary unit of $u$. 
We find double poles with a density of exactly $g$
which means that \eqref{boundHM} and \eqref{boundus} are fully compatible.

In \cite{Hofman:2006xt} it was raised the question about the fate of the bound states
as the coupling decreases. From \eqref{fulleven} we observe that 
there is nothing that prevents 
them from being present all the way to small coupling.
Moreover, as the signs $s^{(n)}$ are stable,
they must appear in the scattering of particles
at weak coupling. This does not necessarily represent a problem for
a smooth interpolation to gauge theory though: 
Firstly, the infinite genus of the function 
leads to various inequivalent weak-coupling limits
(this problem is discussed in the next paragraph, although 
for strong coupling instead of weak coupling). 
Most of the poles could be invisible, their influence being 
however still present in the perturbative series. 
Secondly, it is not clear which configuration of signs $s^{(k)}$
connects to gauge theory magnons in the first place: 
the ``elementary'' choice, the
``giant'' choice or an altogether different choice.
Finally, we do not understand the analytic structure 
of homogeneous solution, see section \ref{sec:homo}. 
Including the right homogeneous solution could, 
in principle, cure the disagreement between
the asymptotic phases for both models.

Note that the phase $\theta\supup{giant}_{12}$ in \eqref{fullevenHM}
is defined on an infinite-genus surface. This has the interesting
implication that there exist infinitely many 
inequivalent strong-coupling limits for it, although only the
one considered above is directly associated with the series \eqref{chisum,coeffs}.
These limits can be deformed into each other by going to
finite coupling, changing the sheet, and going
back to infinite coupling. We could for instance take the HM limit $\xp{}\approx 1/\xm{}$
while making sure that 
$\xexp{4n+1}{}\approx 1/\xexp{4n+3}{}\approx \xexp{-4n-1}{}\approx 1/\xexp{-4n-3}{}\approx \xm{}$ 
for $n\geq0$.
By this staggering, adjacent poles and zeros 
will cancel each other in the strong coupling limit 
and only a few singularities
near $n=0$ will remain. The latter will however not
contribute at order $\order{g}$ and the strong
coupling limit becomes simply
\[
\theta\supup{giant}_{12}
=g \bigbrk{p_2\sin(\half p_1)-p_1\sin(\half p_2)}+\order{1/g^0}
=\theta\supup{elem}_{12}+\order{1/g^0} \ .
\]
This particular phase actually has the same leading $\order{g}$ 
contribution as the (unique) HM limit of the elementary phase 
in \eqref{eq:evenanalytic} with all $s^{(n)}=+1$.
Notice that, correspondingly, the HM limit for a particle with $s^{(n)}=+1$
is not compatible with the choice $x^{(n)}\approx x^{(n+1)}$.
Hence the elementary phase is not either directly representable 
by the series \eqref{chisum,coeffs} in this kinematical regime.
Although in the HM regime $\theta\supup{elem}$ and the limit of 
$\theta\supup{giant}$ 
just considered coincide at leading order, the agreement between 
the two expressions will clearly break down at higher orders in $1/g$
and lead to the much richer structure 
of $\theta\supup{giant}$.  

\subsection{Bound State Scattering}

To understand better the additional terms
in the general dressing phase \eqref{fulleven}
it will be instructive to 
consider scattering of bound states \cite{Dorey:2006dq}.
This will reveal that particles with $s^{(n)}\neq +1$ for any $n$
may potentially correspond to some non-minimal bound states.

A bound state can be thought of as composed
from $m$ elementary particles. The particles
are parametrised by $\xpm{1(k)}$, $k=1,\ldots m$, with
\cite{Dorey:2006dq,Roiban:2006gs,Chen:2006gq}
\[\label{boundx}
\xp{1(k)}=\xexp{-m+2k}{1},\qquad
\xm{1(k)}=\xexp{-m+2k-2}{1}\ ,
\]
where the $\xexp{k}{1}$ are parameters
obeying \eqref{xndef}.
In particular, we shall denote the extremal 
parameters which define the multiplet under
the residual symmetry algebra by
\[
\xexpb{-m}{1}=\xm{1(1)},\qquad
\xexpb{+m}{1}=\xp{1(m)} \ .
\]
The total dressing phase for the scattering of two
such bound states is given by
\[\label{eq:bounddress}
\theta_{12}=
\sum_{k=1}^m\sum_{l=1}^n\theta_{1(k)2(l)} \ .
\]
When expressing the phase in 
terms of $\chi$ using \eqref{chi} one finds
that both sums telescope to
\cite{Roiban:2006gs,Chen:2006gq}
\<\label{chibound}
\theta_{12}\eq
+\chi(\xexpb{+m}{1},\xexpb{+n}{2})
-\chi(\xexpb{+m}{1},\xexpb{-n}{2})
-\chi(\xexpb{-m}{1},\xexpb{+n}{2})
+\chi(\xexpb{-m}{1},\xexpb{-n}{2})
\nl
-\chi(\xexpb{+n}{2},\xexpb{+m}{1})
+\chi(\xexpb{-n}{2},\xexpb{+m}{1})
+\chi(\xexpb{+n}{2},\xexpb{-m}{1})
-\chi(\xexpb{-n}{2},\xexpb{-m}{1})\ .
\quad
\>
When we now substitute the 
explicit expression \eqref{chianalytic} leading to the (even part of the)
phase, we find in analogy to \eqref{eq:evenanalytic}
\<\label{evenbound}
\theta\supup{even}_{12}
\eq
-\frac{i}{2}\log \frac{1-1/\xexpb{-m}{1}\xexpb{+n}{2}}{1-1/\xexpb{+m}{1}\xexpb{-n}{2}}
\nl
-\frac{i}{2}\sum_{k=1}^{m-1}\log \frac{1-1/\xexp{+m-2k}{1}\xexpb{+n}{2}}{1-1/\xexp{+m-2k}{1}\xexpb{-n}{2}}
-\frac{i}{2}\sum_{l=1}^{n-1}\log \frac{1-1/\xexpb{-m}{1}\xexp{+n-2l}{2}}{1-1/\xexpb{+m}{1}\xexp{+n-2l}{2}}
\nl
+\frac{g}{2}\lrbrk{\frac{1}{\xexpb{+m}{1}}-\frac{1}{\xexpb{-m}{1}}}\log \frac{\xexpb{+n}{2}}{\xexpb{-n}{2}}
-\frac{g}{2}\lrbrk{\frac{1}{\xexpb{+n}{2}}-\frac{1}{\xexpb{-n}{2}}}\log \frac{\xexpb{+m}{1}}{\xexpb{-m}{1}}\,.
\>
Here we have assumed $\xexp{k-m}{}(\xexpb{+m}{1})=\xexp{k+m}{}(\xexpb{-m}{1})=\xexp{k}{1}$
and likewise for $x_2$. This is equivalent to setting all signs defined by the 
analogous to \eqref{matchmap} in this more general case to $+1$.
This answer is indeed 
consistent with substituting
$\theta\supup{elem}$ from \eqref{eq:evenanalytic} in \eqref{eq:bounddress}.

Superficially \eqref{chibound} suggests that all the intermediate $\xexp{k}{1,2}$
do not matter for scattering of bound states.
This is however not true as can be seen from \eqref{evenbound}.
In particular, the expression does make
a distinction between $\xexp{k}{1,2}$ and $1/\xexp{k}{1,2}$.
The infinite genus of the function $\chi\supup{even}$ allows
for the $\xexp{k}{1,2}$ to be reintroduced;
this feature has some interesting consequences:

Let us for instance briefly compare 
to the results in \cite{Roiban:2006gs,Chen:2006gq}
for the total scattering factor for bound states.
When we supplement the above dressing phase $\theta\supup{even}_{12}$
with the expression in \cite{Roiban:2006gs,Chen:2006gq} 
based on the BDS phase \cite{Beisert:2004hm},
we find
\<\label{eq:fullboundscat}
S\supup{BDS+even}_{12}
\eq
 \frac{\xexpb{+m}{1}-\xexpb{-n}{2}}{\xexpb{-m}{1}-\xexpb{+n}{2}}
\prod_{k=1}^{m-1}\frac{\xexp{+m-2k}{1}-\xexpb{-n}{2}}{\xexp{+m-2k}{1}-\xexpb{+n}{2}}
\prod_{l=1}^{n-1}\frac{\xexpb{+m}{1}-\xexp{+n-2l}{2}}{\xexpb{-m}{1}-\xexp{+n-2l}{2}}
\nlnum\nonumber
\times\exp\left[ig\lrbrk{\frac{1}{\xexpb{+m}{1}}-\frac{1}{\xexpb{-m}{1}}}\log \frac{\xexpb{+n}{2}}{\xexp{-n}{2}}
-ig\lrbrk{\frac{1}{\xexpb{+n}{2}}-\frac{1}{\xexpb{-n}{2}}}\log \frac{\xexpb{+m}{1}}{\xexpb{-m}{1}}
\right]\,.
\>
One can observe that the double poles 
found in \cite{Roiban:2006gs,Chen:2006gq}
may now split up into two separate poles depending
on the intermediate $\xexp{k}{1,2}$.
For example, the equality of
$\xexpb{-m}{1}=\xexp{+n-2l}{2}$
does not necessarily imply
$\xexp{-m+2l}{1}=\xexpb{+n}{2}$
and the corresponding poles in the third 
and second terms might not overlap.

It is also curious to see that the 
terms involving intermediate parameters of one bound state, e.g.~$\xexp{k}{1}$,
depend on the other bound state only via the extremal parameters, e.g.~$\xexpb{\pm n}{2}$.
This is in fact the same pattern as for
the additional terms in \eqref{fulleven},
so let us write the analogous terms for this case explicitly
\<\label{eq:switchers}
\delta\theta\supup{even}_{12}
\eq
-\frac{i}{4} \! \sum^\infty_{k=-\infty} \sign(2k-1) \,
\frac{1-s^{(+m-2k)}_1}{2}
\log
\frac{\xexp{+m-2k}{1}-\xexpb{+n}{2}}{\xexp{+m-2k}{1}-\xexpb{-n}{2}}\,
\frac{1/\xexp{+m-2k}{1}-\xexpb{-n}{2}}{1/\xexp{+m-2k}{1}-\xexpb{+n}{2}}
\nl
-\frac{i}{4} \! \sum^\infty_{l=-\infty}\sign(2l-1) \,
\frac{1-s^{(+n-2l)}_2}{2}
\log
\frac{\xexpb{-m}{1}-\xexp{+n-2l}{2}}{\xexpb{+m}{1}-\xexp{+n-2l}{2}}\,
\frac{\xexpb{+m}{1}-1/\xexp{+n-2l}{2}}{\xexpb{-m}{1}-1/\xexp{+n-2l}{2}}  \, ,
\qquad
\>

Interestingly, we can remove the explicit reference
to all the intermediate $\xexp{+m-2k}{1}$ by setting $s^{(+m-2k)}_1=-1$
for all $0<k<m$. In that case, the terms on the first line in \eqref{eq:fullboundscat} 
combine with the ones in \eqref{eq:switchers} to give
\<
S\supup{BDS+even}_{12}
\eq
 \frac{\xexpb{+m}{1}-\xexpb{-n}{2}}{\xexpb{-m}{1}-\xexpb{+n}{2}}
\prod_{k=1}^{m-1}
\sqrt{
\frac{\xexpb{-n}{2}}{\xexpb{+n}{2}}
\frac{u_1-u_2+i(m-2k+n)/2g}{u_1-u_2+i(m-2k-n)/2g}
}
\nl\qquad
\times\prod^{n-1}_{l=1}
\sqrt{
\frac{\xexpb{+m}{1}}{\xexpb{-m}{1}}
\frac{u_1-u_2+i(-n+2l+m)/2g}{u_1-u_2+i(-n+2l-m)/2g}
}
\times\ldots\,. 
\>
Here we have expressed all the intermediate parameters $\xexp{k}{1}$
through $u_1$ and thus through 
the extremal parameters $\xexpb{\pm m}{1}$ 
via $u_1=\xexpb{-m}{1}+1/\xexpb{-m}{1}+im/2g$ only.
The choice of intermediate signs $s^{(+m-2k)}_1=-1$ for $0<k<m$ in bound states 
therefore lowers the genus of the scattering phase. If we also set
$s^{(+m-2k)}_1=+1$ otherwise, we obtain a scattering phase with minimal
genus one, i.e.~it is defined on a complex torus just as the elementary phase. 
Furthermore, most of the square roots in the above
expression appear twice, such that there will
almost be no fractional poles. 
Therefore this appears to be a natural choice for
particles transforming in bigger representations 
of the residual symmetry such as the bound states.

What if we set the exterior sign $s^{(+m-2k)}_1$ for $k<0$ or $k>m$ to $-1$? 
It activates explicit dependence of the phase on $\xexp{+m-2k}{1}$
suggesting that the bound state becomes non-minimal.
This view is reinforced by the fact that there are additional singularities at
$\xexp{+m-2k}{1}$ in the scattering phase which
indicate the presence of some new substructure in the bound state.
However, the new parameter lies outside the range of
the constituent parameters in \eqref{boundx}
so it is not expected to appear in the minimal case.
A possible conclusion would be that
the scattering phase belongs to 
a novel kind of extended bound state.

In any case, it seems that non-trivial signs $s^{(k)}=-1$ play a role 
especially for bound states. 
Along these lines, one might consider the
choice $s^{(k)}=-1$ for all $k$ to correspond to some bound state.
As the giant magnons phase requires precisely this choice, 
one may draw the conclusion that the giant magnons represent 
some extended type of bound state. This agrees with the fact that 
its scattering phase has many more singularities than one might expect
for fundamental particles.  
We however feel that more rigorous investigations are required
to probe the nature of the signs, giant magnons and the proposed 
extended bound states.

We would also like to mention that the
terms in \eqref{eq:switchers} resemble
the monodromies of the one-loop phase $\theta^{(1)}$ in \eqref{thetamono} 
which may appear
as ambiguities in the definition of phase for bound states.
For the choice $\xpm{1}\to\xexp{+m-2k}{1}$
and $\xpm{2}\to\xexpb{\pm n}{2}$ we find the same contributions but
with a prefactor which is four times as large. 
Clearly, more work is needed to fully understand the analytic
structure of the dressing phase, especially in the case of bound states.

\section{Homogeneous Solutions}
\label{sec:homo}

In the previous sections we have discussed a particular solution to
the crossing relation \eqref{crossoddeven}. However, the weak-coupling limit
of this solution apparently does not agree with planar ${\cal N}=4$ Yang-Mills.
Indeed, the series \eqref{chisum, coeffs} can be continued to weak coupling
without
encountering negative powers of $g$. Each $\chi^{(n)}$, with $n>1$, leads
to a weak coupling contribution of order $g^2$ (see App. B), while
the gauge theory phase vanishes at least at order $g^4$ \cite{Beisert:2004hm}.
In addition, $\chi^{(1)}$ \eqref{odd} gives rise to a non-analytical
weak-coupling contribution at order $g^3$. The fact that the phase
we have proposed does not connect with gauge theory
strongly points towards the need for additional pieces with a
different weak-coupling behaviour. If crossing symmetry holds, they must
correspond to homogeneous solutions \eqref{crossoddeven}.
The study of these solutions will be
the subject of this section.

\subsection{General Perturbative Solution}

The form of the strong-coupling coefficients \eqref{coeffs}
suggest that the $c^{(n)}_{r,s}$ may be written
as polynomials in $r$ and $s$ with the degree determined
by the loop order $n$.
It is not difficult to find a general expression for
$c\supup{hom}_{r,s}$ of this form
\[\label{eq:genhom}
c^{\mathrm{hom}}_{r,s}
=
\sum_{m=0}^\infty\sum_{n=m+1}^\infty
b_{m,n}
\bigbrk{1-(-1)^{r+s}}
\bigbrk{(r-1)^{2n+1}(s-1)^{2m+1}-(r-1)^{2m+1}(s-1)^{2n+1}}\ ,
\]
where $b_{m,n}$ are some arbitrary real coefficients
which may depend on the coupling constant $g$.
These contributions sum up to
\[
\chi\supup{hom}(x_1,x_2)=
\sum_{m=0}^\infty\sum_{n=m+1}^\infty
b_{m,n}\bigbrk{\Li_{-2m}(x_1)\Li_{-2n}(x_2)
-\Li_{-2m}(-x_1)\Li_{-2n}(-x_2)}+\ldots\ ,
\]
where the dots represent terms which are symmetric
under the interchange of $x_1$ and $x_2$ and which
consequently drop out in the full phase $\theta_{12}$, cf.~\eqref{chi}.
Using the identity
\[
\Li_{-n}(1/x)+(-1)^{n}\Li_{-n}(x)=-\delta_{n,0}\qquad\mbox{for }n\geq 0\,
\]
it is straightforward to show that $\theta\supup{hom}_{12}+\theta\supup{hom}_{\bar 12}=0$.

Let us now consider the analytic structure of $\theta\supup{hom}$.
Note that $\Li_{-n}(x)$ is a polynomial in $1/(x-1)$ of degree $n$.
Therefore the homogeneous solution corresponding to a single coefficient $b_{m,n}$
has multiple poles at $x_{1,2}=\pm 1$. Thus each non-zero coefficient $b_{m,n}$
gives rise to essential singularities in the phase factor $\sigma\supup{hom}_{12}$.
The only way to get rid of them is by taking either no homogeneous terms,
or infinitely many. Taking infinitely many terms may lift essential singularities,
but usually at the price of branch cuts. However the appearance of branch cuts
has the potential to destroy the homogeneous nature of the solutions.
It is thus rather difficult to figure out suitable non-trivial solutions.

The most trivial solution, with $\theta\supup{hom}=0$, contains a relatively small
number of singularities, and could seem thus a reasonable solution within string theory.
However, since it does not connect with gauge theory at weak coupling
it needs to include additional pieces with a different
weak-coupling behaviour if the AdS/CFT correspondence is correct.
In the following section we will comment on a natural
non-trivial homogeneous solution which could have a chance of
being part of the correct physical answer.

\subsection{Special Solution}

A special perturbative solution of crossing
is given by the odd-$n$ contributions of
\eqref{coeffs} with $n\geq 3$
\[
\theta\supup{hom}=\sum_{n=1}^\infty \theta^{(2n+1)}.
\]
Superficially it may seem that all these are
zero due to the coefficient $\bernoulli_n=0$ for odd $n\geq 3$.
However, the term $\cos(\half \pi n)$ in the denominator also
vanishes for odd $n\geq 3$ and therefore we need to regularise
$c^{(n)}_{r,s}$. A natural extension of the Bernoulli numbers is to use the identity
\[\bernoulli_{n}
=-\frac{2\gammafn(n+1) \cos(\half \pi n)}{(-2\pi)^{n}}\, \zeta(n)
\]
to replace them by the Riemann zeta function.
Then the coefficients $c^{(n)}_{r,s}$
can be seen to be anti-symmetric under the interchange of $r$ and $s$.
Furthermore, they are odd polynomials in $r-1$ as well as in $s-1$.
Therefore the properties of $c^{(n)}_{r,s}$
agree with all the properties of
the general homogeneous solution \eqref{eq:genhom}
and consequently the above $\theta\supup{hom}$ represents a special homogeneous solution.

As this homogeneous solution is a natural extension of the inhomogeneous one,
it could point towards the correct physical answer.
The first contribution of this type
appears at three world sheet loops and reads
\[\label{chioddexp}
\chi^{(3)}(x_1,x_2)=-\frac{x_2+x_1x_2^2+6x_2^3-6x_1x_2^4+x_2^5-3x_1x_2^6}{32g^2(x_1x_2-1)^2(x_2^2-1)^3}
\,\frac{\zeta(3)}{\pi^3}\,.
\]
The higher-loop contributions have a similar form,
however, we do not know the
analytic structure of the sum $\chi\supup{hom}$.
Let us only mention that,
in contrast to the even-$n$ contributions,
the expansion of $c\supup{hom}_{r,s}(g)$ does
not stop at $g^{s-r}$ as in the case of
$c\supup{even}_{r,s}(g)$.
Thus, adding this piece will substantially alter the weak-coupling
behaviour.

\subsection{Rational Solutions}

In the preceding sections we have already seen 
a couple of explicit solutions to the homogeneous crossing equation. 
Let us collect and investigate them briefly here. 
None of these solutions will actually be of the perturbative form proposed 
in section \ref{sec:seriesrep}. 

One of the homogeneous solutions is proportional to
\[
C_1p_2-C_2p_1=
g\lrbrk{\frac{1}{\xp{1}}-\frac{1}{\xm{1}}-\frac{i}{2g}}\log \frac{\xp{2}}{\xm{2}}
-g\lrbrk{\frac{1}{\xp{2}}-\frac{1}{\xm{2}}-\frac{i}{2g}}\log \frac{\xp{1}}{\xm{1}}\,.
\]
In principle, this term could also be written using $\xexp{\pm m}{12}$ 
instead of $\xpm{12}$. Nevertheless the function alters the 
strong-coupling limit substantially unless suppressed by sufficiently many
powers of $1/g$ or if it appears in a suitable linear combination. 
On its own, we can exclude it.

It is also clear that the monodromies of the one-loop contribution
\eqref{thetamono} are homogeneous solutions as well. By themselves
they violate unitarity,%
\footnote{Note that the one-loop solution does not violate unitarity
due to its branch cuts. More explicitly,
the variables $q^\pm$, cf.~\protect\eqref{qvar}, 
will be exchanged. 
The function $\psi(q)$ in \protect\eqref{oddaux} then produces
extra instances of the monodromy terms and unitarity is recovered.}
but they can be symmetrised w.r.t.~unitarity
and one obtains the function $h\supup{odd}$ governing the odd part of the crossing relation,
cf.~\eqref{hoddeven}.
A slight generalisation gives the following homogeneous solution
\<\label{cdd}
-\frac{i}{2}
\log
\frac{\xexp{+m}{1}-\xexp{+n}{2}}{\xexp{+m}{1}-\xexp{-n}{2}}\,
\frac{\xexp{-m}{1}-\xexp{-n}{2}}{\xexp{-m}{1}-\xexp{+n}{2}}\,
\frac{1-1/\xexp{+m}{1}\xexp{-n}{2}}{1-1/\xexp{+m}{1}\xexp{+n}{2}}\,
\frac{1-1/\xexp{-m}{1}\xexp{+n}{2}}{1-1/\xexp{-m}{1}\xexp{-n}{2}}\,,
\>
Note that this expression is somewhat reminiscent of the CDD poles \cite{Castillejo:1955ed} 
for ordinary S-matrices; here the position of the poles is determined by
the two integer parameters $m,n$.
The transformation of this solution under parity symmetry
depends on how the $\xexp{k}{}$ transform, i.e.~to 
$-\xexp{-k}{}$ or to $-1/\xexp{-k}{}$. To preserve parity,
one of the two $\xexp{k}{}$ needs to transform
to $-\xexp{-k}{}$ and the other one to $-1/\xexp{-k}{}$.
In the analytic expression for the even part of the phase
\eqref{fulleven} this was the case:
The variable $\xpm{}$ transforms without inverse
while $\xexp{2n-1}{}$ transforms with inverse if
$s^{(2n-1)}=-1$. If the sign was $s^{(2n-1)}=+1$ instead,
the variable $\xexp{2n-1}{}$ would transform without inverse,
but also the homogeneous term would be absent due to the prefactor. 
Similarly, the above term \eqref{cdd} should be activated
only if it does not violate parity, i.e.~if 
$s^{(\pm m)}_1=-s^{(\pm n)}_2$.

\section{Conclusions and Outlook}
\label{sec:concl}

In this paper we have constructed a dressing phase factor for the world sheet 
scattering matrix of type IIB string theory on $AdS_5 \times S^5$. 
The general expression that we propose solves the condition 
imposed by crossing symmetry on the dressing factor \cite{Janik:2006dc}, 
and admits an expansion in the strong coupling regime.
The main result of the paper is a proposal for the coefficients governing
this series. The coefficients provide an explicit all-loop expansion of the
dressing phase factor, and contain the leading order term \cite{Arutyunov:2004vx} as well as the
first quantum correction \cite{Beisert:2005cw,Hernandez:2006tk,Freyhult:2006vr}. A direct 
two-loop test of this proposal would be highly desirable and it might even be feasible.

The structure of the perturbative series is not straight-forward because it does
not converge properly. In order to support the proposed expansion, we have shown
how the coefficients satisfy the different pieces in the crossing relation. In
particular, in order to satisfy the odd piece of the crossing relation it suffices
to consider the one-loop contribution to the phase.
This was already suggested in \cite{Beisert:2006zy} and the present work contains a proof
of the statement. Moreover, we specify clearly how the
antipode map must act in order to obey the correct crossing relation.
The even piece of the crossing relation is satisfied when including the even terms in the loop
expansion of the phase. 

In addition, we have found an analytic expression for the
resummed series. The strong-coupling limit of the perturbative series agrees
with this analytic expression. 
Furthermore, the analytic form for the resummed
series allows an analysis of the spectrum of bound states of giant magnons. Bound
states arise as poles of the scattering matrix, and have been found for giant
magnons in \cite{Hofman:2006xt}. We have identified these bound states of giant magnons
using the analytic form of the dressing phase. What complicates the discussion is 
that our analytic expression for the phase involves an arbitrary or even an infinite number of 
branch points. Specially the kinematical space for giant magnons becomes
an infinite-genus surface. 
In contrast, there is also a minimal particle for which the phase 
merely requires a genus-one surface. In order to probe the structure of the
dressing phase, we have furthermore considered scattering of bound states. This
seems to point out toward the possibility that the giant magnon states of \cite{Hofman:2006xt}
are not elementary, but rather composites of some minimal particles. 
A better understanding of this issue as well as the analytic structure of the 
phase in general clearly deserves further study.

We have also presented an abridged study of homogeneous solutions to the crossing
relation. We know only few physical constraints on the homogeneous piece of the
crossing condition, and thus most of the homogeneous solutions 
can be introduced arbitrarily. 
A careful look at them is worthwhile because they might be at the root of a
discrepancy between our proposed phase and gauge theory: The weak coupling limit
of the analytic phase disagrees with gauge theory, as opposed to the
agreement in the strong-coupling regime with string theory. 
In particular, our perturbative phase includes homogeneous pieces which we
were not able to sum up to an analytic expression.  
These homogeneous solutions most likely change the weak-coupling behaviour, 
and open the possibility for a cure
of the disagreement in the gauge theory limit. 
Homogeneous solutions could also clarify the nature, or even the existence, of the fractional 
singularities in the scattering matrix for the general parity-invariant dressing phase 
that we have constructed. 
Further research on
homogeneous solutions to the crossing condition could clarify the existence of
a smooth interpolating function from the string to the gauge theory scattering
matrices. A three-loop string theory calculation would verify or disprove the first 
homogeneous piece in our perturbative phase, but unfortunately this
is most likely beyond the current computational abilities.

Finally let us note that in the present work we have not considered particles 
whose energy and momentum scale as $\lambda^{\pm 1/4}$ (GKP regime). An investigation 
of this kinematical regime at strong coupling may be particularly interesting because 
it contains some of the special points in the phase. Furthermore, it seems that the 
structure of the perturbation series should be changed which possibly enables different tests 
of our proposal. Another interesting class of states are the `antiferromagnetic' states
\cite{Park:2005kt,Rej:2005qt,Zarembo:2005ur,Roiban:2006jt,Feverati:2006tg,Beccaria:2006td}
whose study in the current framework might lead to further insight.


\subsection*{Acknowledgements}

We are grateful to A.~Donini, N.~Dorey, M.~Garc{\'\i}a P\'erez, A.~Gonz\'alez-Arroyo,
R.~Janik, K.~Landsteiner, J.~Maldacena
and especially M.~Staudacher for clarifying discussions and comments.
The work of N.B.~is supported in part by the U.S.~National Science
Foundation Grant No.~PHY02-43680. Any opinions, findings and conclusions or
recommendations expressed in this material are those of the authors and do not
necessarily reflect the views of the National Science Foundation. 
The work of E.L.~is supported by a Ram\'on y Cajal contract of MCYT and in part
by the Spanish DGI under contracts FPA2003-02877 and FPA2003-04597.

\appendix

\section{Strong and Weak Coupling}
\label{sec:strongweak}

In this appendix we collect useful expressions to
parametrise the kinematics of particles
in the strong and weak-coupling limits.

\subsection{Strong Coupling}
\label{sec:stronglimit}

Let us start with strong coupling, $g\to\infty$.
In this case there are four interesting and distinct regions for the kinematic
space.
This can most easily be seen by considering relation \eqref{def}.
For large values of $g$ we can solve constraint \eqref{def} by setting either
$\xp{}\approx\xm{}$ or $\xp{}\approx1/\xm{}$ or $\xp{}\approx \xm{}\approx \pm 1$.
These regions remind curiously of the different types of particles considered
in \cite{Mann:2005ab}.%
\footnote{
In general, it would be interesting to recover 
our proposal from a covariant framework 
as in \protect\cite{Mann:2005ab,Rej:2005qt,Gromov:2006dh,Gromov:2006cq}
which has proved to work at least to the leading order.}

Note that we will use a relativistic rapidity variable $\vartheta$
(different in all four cases) to parametrise the momenta of particles.
The energy and momentum must be periodic
in shifts of $\vartheta$ by $2\pi i$
and real for real values of $\vartheta$.
Furthermore, increasing $\vartheta$ should increase the momentum.

\paragraph{Metsaev-Tseytlin Regime.}

The first class of solutions to \eqref{def} is
\[\label{MT}
\xpm{}=\coth(\half\vartheta)\lrbrk{1\pm \frac{i}{4g}\sinh\vartheta}+\order{1/g^2} \ ,
\]
The resulting momentum, energy and $u$-parameter are given by
\[
p=\frac{\sinh\vartheta}{2g}+\order{1/g^2}\ ,
\quad
C=\half \cosh\vartheta+\order{1/g}\ ,
\quad
u=2\coth\vartheta+\order{1/g^2}\ .
\]
Most importantly, the momentum $p=\order{1/g}$ is very small.
This combination corresponds to particles which behave
like elementary excitations \cite{Metsaev:1998it} in the plane-wave limit,
see \cite{Berenstein:2002jq}.
They also serve as the quantum constituents for the Frolov-Tseytlin
spinning string solutions \cite{Frolov:2002av,Frolov:2003qc}, see \cite{Beisert:2003xu,Beisert:2003ea}.

\paragraph{Hofman-Maldacena Regime.}

The second solution is 
\[\label{HM}
\xpm{}=-\tanh\vartheta\pm \frac{i}{\cosh\vartheta}+\order{1/g}=e^{\pm ip/2}+\order{1/g}
\]
The resulting momentum, energy and $u$-parameter are given by
\[
p=\pi+2\arctan\sinh\vartheta+\order{1/g}\ ,
\quad
C=\frac{2g}{\cosh\vartheta}+\order{1/g^0}\ ,
\quad
u=-2\tanh\vartheta+\order{1/g}\ .
\]
In this case the momentum is finite, $p=\order{1/g^0}$,
and its range is given by $0<p<2\pi$.
This is the limit investigated by Hofman and Maldacena \cite{Hofman:2006xt}
with ``giant'' magnons as excitations.
It is also the region of the spinning string solutions first found in \cite{Gubser:2002tv}.

\paragraph{Gubser-Klebanov-Polyakov Regime.}

The last two solutions involve square roots of $g$,
\[
\xpm{}=s+\frac{se^{-s\vartheta}\pm ie^{s\vartheta}}{2\sqrt{g}}+\order{1/g}\ ,
\]
where $s=\pm 1$ distinguishes the two solutions.
In this case we find
\[
p=s\,\frac{e^{s\vartheta}}{\sqrt{g}}+\order{1/g^{3/2}}\ ,
\quad
C=\sqrt{g}\,e^{s\vartheta}+\order{1/g^{1/2}}\ ,
\quad
u=2s-\frac{\sinh(2\vartheta)}{2g}+\order{1/g^2}\ .
\]
Now the momentum is small, $p=\order{1/g^{1/2}}$, but not
as small as in the above case.
This region comprises the states
whose energy scales as
$\sqrt{g}\sim\sqrt[4]{\lambda}$
studied by Gubser, Klebanov and Polyakov \cite{Gubser:1998bc}.
As found in \cite{Arutyunov:2004vx}, the particles split
up in right movers with $s=+1$ and left movers with $s=-1$.

\subsection{Weak Coupling}
\label{sec:weaklimit}

For weak coupling $g\to 0$, we find two distinct regions
where the particle momenta are real.
We shall use the momentum $p$ as the fundamental parameter.

\paragraph{Magnons.}

The standard magnons correspond to the solution
\[
\xpm{}=\frac{e^{\pm ip/2}}{2g \sin(\half p)}+\order{g}
\]
The resulting energy and $u$-parameter are given by
\[
C=\half+4g^2\sin^2(\half p)+\order{g^4}\ ,
\quad
u=\frac{\cot(\half p)}{2g}+\order{g}\ .
\]
%

\paragraph{Holes.}

The other relevant solution comprises magnon-holes with
\[
\xpm{}=-\frac{2g \sin(\half p)}{e^{\mp ip/2}}+\order{g^3}
\]
Their energy and $u$-parameter read
\[
C=-\half-4g^2\sin^2(\half p)+\order{g^4}\ ,
\quad
u=-\frac{\cot(\half p)}{2g}+\order{g}\ .
\]
%

\section{Weak-Coupling Expansion}
\label{sec:WeakBorel}

At weak coupling and for standard magnons, the $\xpm{}$ variables
scale as $\xpm{}\sim 1/g$.
The combination $c^{(n)}_{r,s}q_r(\xpm{1})q_s(\xpm{2})$ consequently
scales as $g^{r+s-n-1}$. The lowest-order terms
in $\chi\supup{even}$ therefore appear
for $n$ as large as possible, i.e.~for $n=s-r+1$.
This means that for fixed $r,s$ the lowest order is $g^{2r-2}$
and globally it is $\order{g^2}$ with terms of $r=2$ and arbitrary odd $s$
contributing
\[
\chi\supup{weak-LO}(x_1,x_2)=
-\sum_{s=3}^\infty
\frac{c^{(s+1)}_{2,s}}{(s-1) x_1 x_2^{s-1}}=
\sum_{n=2}^\infty
\frac{i^{n}\bernoulli_{n}}{2ng^{n-1} x_1 x_2^{n}}\ .
\]
This series is not well-defined due to the
asymptotics of the Bernoulli numbers $\bernoulli_n\sim n!$ 
A standard procedure in field theory is to Borel
sum the series. This can be done in the present case.
Let us first define a function
\[\label{eq:Hser}
H(x)=
\log (igx)
-\sum_{n=2}^\infty
\frac{i^{n}\bernoulli_{n}}{n\,g^{n} x^{n}}
\]
such that
\[
\chi\supup{weak-LO}(x_1,x_2)=
\frac{g}{2x_1}\,\log(ig x_2)
-\frac{g}{2x_1}\,H(x_2)\ .
\]
First of all we perform an inverse Laplace transformation
assuming that $igx$ has a positive real part.
Then the sum can be performed due to improved convergence
\<
H(x)\eq
\log (igx)
-\int_0^\infty dt \exp(-igxt)
\sum_{n=2}^\infty
\frac{\bernoulli_{n}}{n!}\,t^{n-1}
\nln\eq
\log (igx)
-\int_0^\infty dt \exp(-igxt)
\lrbrk{\frac{1}{2}\coth(t/2)-\frac{1}{t}}\ .
\>
Finally, we perform the Laplace transformation integral
to recover an analytic function
\[
H(x)=\digamma(igx)+\frac{1}{2igx}
\label{eq:BorelH}
\]
involving the digamma function $\digamma(z)=\partial_z\log\gammafn(z)$.
The expansion of this function for large and \emph{positive} $igx$
in fact agrees with the series \eqref{eq:Hser}.
For large negative $igx$ the function oscillates strongly
which explains the divergence of the series.

We can now convert $\chi\supup{weak-LO}$ to the dressing phase
using \eqref{chi}. It is curious to see that the digamma function
appears in the combination $\digamma(ig\xp{})-\digamma(ig\xm{})$
and that at weak coupling $ig\xm{}-ig\xp{}\approx 1$.
Together, the two facts lead to a large cancellation between the
digamma functions and one is left with the simple term $i/g\xp{}$.
In any case, the resulting phase is non-zero at $\order{g^2}$.
It is therefore clear that this result does not agree with
planar gauge theory for which the phase is zero at
least at $\order{g^4}$ \cite{Beisert:2004hm,Beisert:2003ys,Eden:2004ua}.

Nevertheless, it may be important to find the higher-order
corrections at weak coupling.
It turns out that these can
also be written in terms of the function $H(x)$ and its derivatives.
We find that the correct expansion of $\chi\supup{even}$ is encoded in the function
\[\label{eq:BorelT}
T(x_1,x_2,t)=
\sum_{r=2}^\infty
\sum_{m=0}^\infty
\sum_{k=0}^m
\frac{g\,(-1)^{r-1}t^{r+2m-k-2}}{2(m-k)!\,k!\,(r+m-k-1)! \,x_1^{r-1} x_2^k}
\]
which acts on $H(x_2)$ as follows
\<\label{eq:TonH}
\chi\supup{even}(x_1,x_2)\eq\frac{g}{2x_1}\log(i g x_2)+
\mathopen{:}T(x_1,x_2,\partial/\partial x_2)\mathclose{:}\,H(x_2)
\nlnum\nonumber
+\frac{g}{2}\lrbrk{-\frac{1}{x_1}-\frac{1}{x_2}}
+\frac{g}{2}\lrbrk{-x_1-x_2+\frac{1}{x_1}+\frac{1}{x_2}}
\log\lrbrk{1-\frac{1}{x_1x_2}}.
\>
The bracket $:\ldots:$ implies
normal ordering between the variable $x_2$ and its derivative operator
$\partial/\partial x_2$.
The equivalence between \eqref{eq:TonH,eq:BorelT,eq:BorelH}
and \eqref{chisum,coeffs}
can be checked straightforwardly using the identity
\[
\sum_{k=0}^m
\frac{(-1)^{k-1}(n+r+2m-k-3)!}{(m-k)!\,k!\,(r+m-k-1)!}
=
\frac{\gammafn(m+n-1)\,\gammafn(m+n+r-2)}{\gammafn(n-1)\,\gammafn(m+1)\,\gammafn(m+r)}
\]
and substituting $m=(s-r+1-n)/2$ in the final expression.%
\footnote{The terms with $n=0$ are reproduced by the expression
only up to terms which are symmetric under the interchange of
$x_1$ and $x_2$. These are cancelled by the terms
on the second line in \protect\eqref{eq:TonH}.}

Let us remark that the function $T$ can be summed up using an
integral of the Bessel function $I_0$
\[
T(x_1,x_2,t)=-\frac{g}{2x_1}\exp\lrbrk{\frac{t}{x_2}-\frac{t}{x_1}}
\int_0^1 dq  \exp\lrbrk{\frac{t}{x_1}\,q} I_0\bigbrk{2t\sqrt{q}}\  .
\]
The expansion of $T$ for large $x_1,x_2$ and small $t$ reads
\[
T(x_1,x_2,t)=
-\frac{g}{2x_1}
+\frac{gt(x_2-tx_1x_2-2x_1)}{4x_1^2x_2}
+\ldots
\]
in agreement with the series \eqref{eq:BorelT}.

It is now not difficult to compare also the weak coupling expansion
of the analytic expression \eqref{chianalytic}
with the above procedure.
We find perfect agreement with the expansion of the function
\<\label{eq:evenleft}
\chi\supup{even-left}(x_1,x_2)
\eq{}
\lim_{N\to\infty}\left[
\frac{g}{2x_1}\log \frac{igx_2}{N}
+\frac{i}{2}\sum^N_{n=1}\log \lrbrk{1-\frac{1}{x^{}_1\xexp{-2n}{2}}}
\right]
\nlnum\nonumber
+\frac{g}{2}\lrbrk{-\frac{1}{x_1}-\frac{1}{x_2}}
+\frac{g}{2}\lrbrk{\frac{i}{2g}-x_1-x_2+\frac{1}{x_1}+\frac{1}{x_2}}\log\lrbrk{1-\frac{1}{x_1x_2}}\ .
\>
at the leading six orders. The expression \eqref{eq:evenleft} does not literally agree 
with \eqref{chianalytic}, but only after symmetrising as follows
\[
\chi\supup{even}(x_1,x_2)=
\half\chi\supup{even-left}(x_1,x_2)
-\half\chi\supup{even-left}(-x_1,-x_2) \ .
\]
The reason for this additional step in comparing can be explained as follows:
The exact expression has an essential singularity at $x_2=\infty$
due to accumulation of singularities. 
Therefore the power series around $x_2=\infty$ could possibly not converge.
In performing the above Borel summation and Laplace transform
we specified that the real part of $igx_2$ is positive.
Effectively, this regularised the resummed expression such
that the singularities approach $x_2=\infty$ with
negative $igx_2$. If we had chosen to use negative
$igx_2$ in the Laplace transform, the resummed
expression would have the singularities approaching
$x_2=\infty$ with positive $igx_2$.
In other words the resummed expression
is ambiguous and in \eqref{chianalytic}
we chose to present the symmetrised
expression which has manifest parity invariance.
This matching provides further evidence for the agreement between
the proposed coefficients \eqref{coeffs} and the
proposed analytic expression \eqref{chianalytic}.


\bibliographystyle{nb}
\bibliography{stringphase}

\end{document}